\pdfoutput=1

\documentclass[10pt,journal,compsoc]{IEEEtran}

\ifCLASSOPTIONcompsoc
  \usepackage[nocompress]{cite}
\else
  \usepackage{cite}
\fi

\ifCLASSINFOpdf
   \usepackage[pdftex]{graphicx}
\else

\fi

\usepackage{balance}  
\usepackage{tabularx}
\usepackage{times}    
\usepackage{url}      

\usepackage{ulem} 

\usepackage{verbatim} 

\usepackage{multirow}

\usepackage{color}
\newcommand{\textred}[1]{\textcolor{red}{#1}}
\ifx\noeditingmarks\undefined
   \newcommand{\pgwrapper}[2]{\textred{#1 #2}}
\else
   \newcommand{\pgwrapper}[2]{}
\fi

\usepackage[lined,boxed,commentsnumbered,linesnumbered]{algorithm2e}

\usepackage{balance}

\begin{document}

\title{Transforming Speed Sequences into Road Rays on the Map with Elastic Pathing}

\author{Xianyi Gao, Bernhard Firner, Shridatt Sugrim,
Victor Kaiser-Pendergrast, Yulong Yang, Janne Lindqvist \\ Rutgers University}


\IEEEtitleabstractindextext{
\begin{abstract}
Advances in technology have provided ways to monitor and measure driving behavior. Recently, this technology has been applied to usage-based automotive insurance policies that offer reduced insurance premiums to policy holders who opt-in to automotive monitoring. Several companies claim to measure only speed data, which they further claim preserves privacy. However, we have developed an algorithm - elastic pathing - that successfully tracks drivers' locations from speed data. The algorithm tracks drivers by assuming a start position, such as the driver's home address (which is typically known to insurance companies), and then estimates the possible routes by fitting the speed data to map data. To demonstrate the algorithm's real-world applicability, we evaluated its performance with driving datasets from central New Jersey and Seattle, Washington, representing suburban and urban areas. We are able to estimate destinations with error within 250 meters for 17\% of the traces and within 500 meters for 24\% of the traces in the New Jersey dataset, and with error within 250 and 500 meters for 15.5\% and 27.5\% of the traces, respectively, in the Seattle dataset. Our work shows that these insurance schemes enable a substantial breach of privacy. 

\end{abstract}

}

\maketitle

\IEEEdisplaynontitleabstractindextext

\IEEEpeerreviewmaketitle

\IEEEraisesectionheading{\section{Introduction}\label{sec:introduction}}
\IEEEPARstart{T}{echnological} advances have created ways to observe driving behaviors using odometer readings and in-vehicle telecommunication devices. Taking advantage of this technology, some US-based automotive insurance
companies~\cite{allstatedrivewise,gmacinsurance,progressivesnapshot,statefarmindrive} offer consumers the ability to opt-in to the usage-based insurance (UBI) for reduced premiums by allowing companies to monitor their driving behavior. These policies are different from traditional insurance policies, which use record of past driving violations to differentiate between safe and aggressive drivers. The UBI policy applies devices to monitor directly while people are driving and provide insurance companies with real-time driving data.

Although the UBI policy has advantages for both insurers and consumers (e.g.~insurers monitor consumers' driving behaviors as a way to encourage safe driving, and consumers get lower premiums), 
the privacy concern of collecting driving behavior data should not be neglected because this is an always on activity. Some of these monitoring devices are even based
upon GPS information and offer no privacy protection, such as OnStar~\cite{OnStar}. Many insurance companies are aware of the privacy issue and only use devices that collect time-stamped speed data, making the claim that it is privacy-preserving. However, it is still possible to deduce location data from speed traces. This represents a huge breach of privacy for drivers that have these types of insurance policies, as we have shown in our prior work~\cite{Gao:2014:EPY:2632048.2632077}.

\begin{figure}[!t]
\centering
\includegraphics[width=7.5cm]{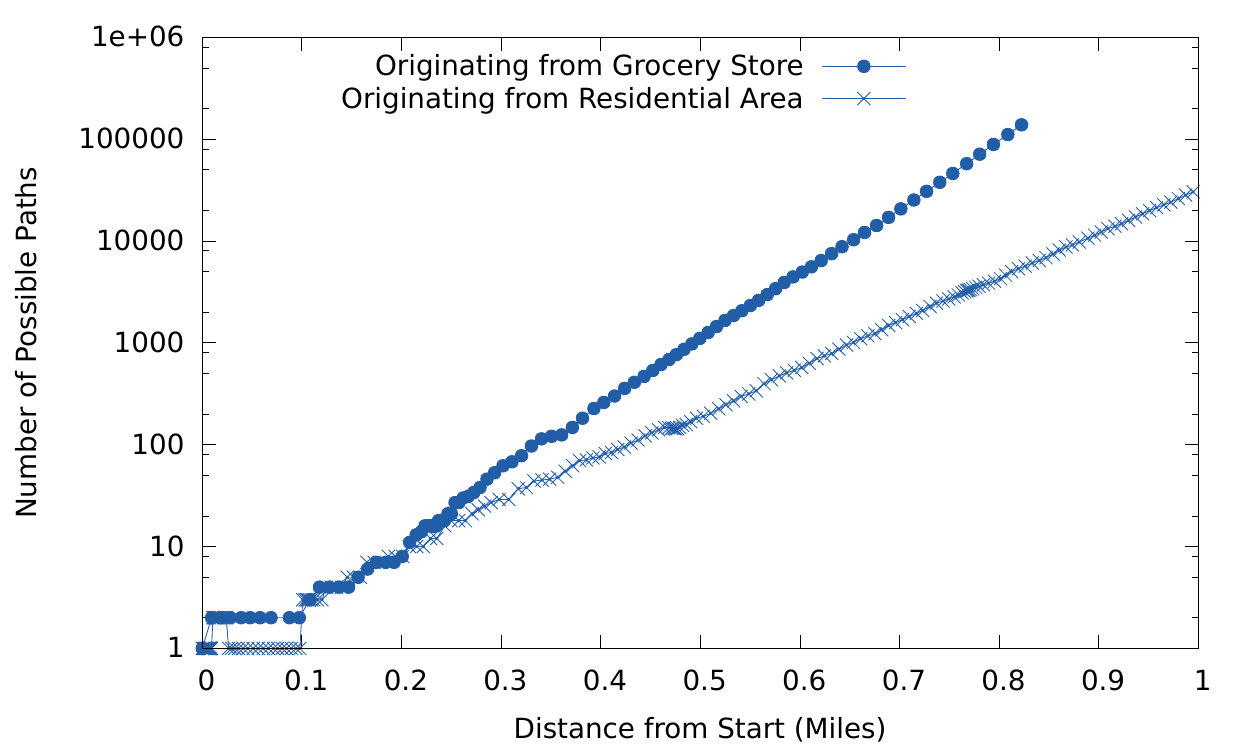}
\caption{Number of paths within a given distance of a starting location. 
The pattern of growth on a log-linear scale shows exponential increments.
Locations central to major transportation routes, such as a grocery store, will see a higher increase
in paths compared to a residential location which is more likely to lead to a dead end.
}
\label{fig:path_growth}
\end{figure}

Tracking drivers and estimating their trip destinations are very challenging with only a starting location and the time-stamped speed data. It is not obvious that these data are sufficient to reproduce an exact driving path. Without information about driving directions, there may be multiple alternative paths that match with the speed data. Blindly exploring each path in a given map is not feasible, since the number of possible path is huge even with a very short driving distance. Figure~\ref{fig:path_growth} shows the growth of
possible paths within a distance of just one mile from a starting location. Within that one mile, there are over 100,000 possible paths the driver could have taken when the trip starts from a grocery store, and over 10,000 paths from a residential area.

Our elastic pathing algorithm shows that drivers can be tracked by merely collecting their speed data and knowing the starting location which is usually their home address that insurance companies have. 
We evaluate our algorithm with datasets from New Jersey and Seattle, Washington, representing suburban and urban areas. Our algorithm estimates destinations with error within 250 meters for 17\% traces and within 500 meters for 24\% traces in the New Jersey dataset (254 traces). For the Seattle dataset (691 traces), we similarly estimate destinations with errors that are within
250 and 500 meters for 15.5\% and 27.5\% of the traces respectively. Our work is important given that speed data is not considered sensitive data. This is why insurance companies claimed that collecting speed data is privacy-preserving. Because of this, it is not likely that this data will be treated as sensitive. The data may eventually be obtained by antagonists who do know how to process and obtain location information. Our intention was to highlight this type of data collection and alarm both customers and insurance providers the amount of location information that may be estimated from the speed data.

Our paper makes following contributions:
\begin{enumerate}
\item We present the definite version of our elastic pathing algorithm which estimates location traces from speed data and starting locations.
\item We built a tool to visualize how our algorithm tracks drivers. Our tool, utilizing Google Maps APIs, can be applied to any application having driving traces.
\item We tested our algorithm on real world traces to show how the data collected by many insurance companies is not privacy-preserving despite their claims. 
\item We explored the effect of adapting OpenStreetMap routing into our algorithm and showed it does not benefit our algorithm but can be potentially useful on estimating routes for a subset of drivers. 
\item 
We analyzed how information about speed limits would affect our estimation accuracy in two distinct driving environments: urban and suburban areas.
\item We collected a new dataset of 30 drivers repeating the same driving routes.
\item We used this new dataset to explore how our algorithm performs for various drivers and driving behaviors.
\item We analyzed traces with low estimation accuracy and summarized factors that may mislead our algorithm. 
\end{enumerate}

\section{Related Work}
\label{sec:related}
Location traces contain a great deal of behavioral information that
people consider private~\cite{Brush:2010:EEU:1864349.1864381,ludford,patil,tsai}, which
has encouraged a large body of work to concentrate on protecting, anonymizing, or obfuscating~\cite{Popa:2011:PAL:2046707.2046781,Gruteser:2003:AUL:1066116.1189037,Krumm:2009:RDT:1560004.1560010,Brush:2010:EEU:1864349.1864381,krumm:inference,Zhou:2013:ILD:2508859.2516661} location traces. Krumm has written an overview of computational location privacy techniques~\cite{Krumm:2009:SCL:1569346.1569363}, and Zang and Bolot have
recently questioned the possibility of releasing privacy-preserving cell phone records while still maintaining
research utility in those records~\cite{Zang:2011:ALD:2030613.2030630}.  

Based on analysis of location data, 
researchers have shown
that the nature of individual mobility patterns is bounded, and that people visit only a few locations
most of the time (e.g.~just two~\cite{barabasi:individual,eagle_Pentland_2009,Golle:2009:AHL:1560004.1560038}). In most cases, there
is high level of predictability for future and current locations
(e.g.~\cite{barabasi:limitspredictability,Farrahi:2011} -- most trivially, ``70\% of the time the
most visited location coincides with the user's actual
location''~\cite{barabasi:limitspredictability}). Mobility patterns can also be used to
predict social network ties (e.g.~\cite{Wang:2011:HMS:2020408.2020581}). In a more applied domain,
GPS traces have been used to learn transportation modes~\cite{Zheng:2010:UTM:1658373.1658374}, to predict routes~\cite{Ziebart:2008:NLC:1409635.1409678}, to predict family routines (e.g.~picking up
or dropping off children)~\cite{Davidoff:2011:LPP:1978942.1979119}, to distinguish when people are home or away~\cite{krumm:learning}, and to recommend friends and
places~\cite{Zheng:2011:RFL:1921591.1921596}. However, none of the above work can be used to discover
locations based solely on speed data and a driver's starting location.

The world's first navigating system from Etak in 1985 showed how well a car could be located without using GPS~\cite{1623458}. The navigating system used a map database, a dead reckoning system with motion sensors, and a magnetic compass to track the driving direction, distance, and ultimately the real time location. This system cannot be simply applied to our work as we do not have the information about driving direction. Other related work in the field of dead reckoning also suggests that speedometer traces should have some
level of information to extract. Dead reckoning works by using speed or movement data starting from
a known location to deduce a movement path.
Dead reckoning has been previously used for map building with mobile
robots~\cite{elastic_correction} or as an addition to
Global Navigation Satellite Systems (GNSS) such as GPS~\cite{kalmann_integration,zhao2003extended}. These systems, however, require very accurate measurements of speed and acceleration, and utilize very precise gyroscopes to determine changes in direction. 
When supplementing GNSS data, dead reckoning was only used when GNSS
information is unavailable, such as when a vehicle passes through a tunnel.

In the specific area of usage-based insurance, the
privacy concerns of these insurance programs have been studied before by Troncoso et al.~\cite{Troncoso:2011:PPP:2015838.2015879}. 
However, this work dates back to schemes which would send raw location (GPS) coordinates to either insurance providers or brokers. Troncoso et al.~proposed
a cryptographic scheme PriPAYD to address the problem~\cite{Troncoso:2011:PPP:2015838.2015879}. Our work shows that speedometer-based solutions, which
were not considered by Troncoso, are not privacy-preserving either.

Technically similar to our work are side-channel attacks using accelerometer data from smartphones to infer user location~\cite{Aviv:2012:PAS:2420950.2420957}.
Projects such as
ACComplice~\cite{accomplice} and AutoWitness~\cite{Guha:2010:ALT:1869983.1869988} have used
accelerometers and gyroscopes for localization of drivers. However, the information from the
smartphone can be used to detect when turns occur. In contrast, we have only a time series
of speed data available. It does not indicate if any turn is taken.

Finally, the most closely related work to our own is a project to infer trip destinations from driving habits data by Dewri et al.~\cite{dewri2013inferring}. Their work used time, speed, and driving distance data to estimate destinations.
Their algorithm is based on depth-first search (DFS) to explore all the paths within the range. After generating candidate paths, post-processing is needed to calculate the actual ranking of paths. The average time complexity is equivalent to the DFS exploration plus the postprocessing, which increases exponentially with the length of speed data. The algorithm also needs to store all the exploring paths within the range. In contrast, our algorithm has much less time and space overhead because of the use of error metrics and prioritizing the search of best paths (see \textbf{Space and Time Complexity} under Section 3.3).

Dewri et.~al.~tested the algorithm with only 30 trips in Denver, Colorado area and mentioned the estimated destinations are close to the actual one (within 0.5 mile)~\cite{dewri2013inferring}. However, each of the 30 trips was pre-processed to ``remove data points that may correspond to driving in traffic condition''~\cite{dewri2013inferring}. The algorithm also assumes that ``all left turns happen at a speed of 15 mph and all right turns happen at 10 mph''~\cite{dewri2013inferring}. It is unclear how this assumption stands for different drivers. The algorithm has not been tested in a large set of non-ideal driving routes with a collection of different drivers and driving habits. In contrast, our elastic pathing algorithm estimated about 37\% of all traces within 0.5 miles using datasets with nearly one thousand daily driving traces. Our algorithm works in realistic driving conditions without manual pre-processing or pre-selecting the collected driving traces.

Another advantage of our elastic pathing algorithm over Dewri's algorithm is the applicability to real-time tracking, since our search algorithm is only based on the current and past speed samples. No post-processing or information of future speed samples are needed to estimate the current location (see Section 3 for algorithm design).

\section{Algorithm Design}
In this section, we describe the algorithm we use to recreate a person's driving path when the time-stamped speed data and the starting location are given. We start with discussing the basic requirements for a generic algorithm and continue by presenting our elastic pathing algorithm. 
In the end, we demonstrate a tool to visualize how our algorithm tracks driving routes.

\subsection{Requirements for a Generic Algorithm}
To solve this path recreation problem, any generic algorithm needs to consider: resources, restrictions, and scoring metric for ranking. Resources are
all of the contextual data that will help us deduce a path from speed data. For example, a map is a useful resource to apply.  Restrictions are physical or behavioral limitations to any automotive driving behavior. Finally, a scoring metric is needed to choose the best path out of several possible candidates. 

\textbf{Applying Available Resources:} First, we would need the information about how different paths are connected in the area. OpenStreetMap (OSM) is a great option for this purpose. A path (or a way~\cite{OSMway}) in OSM is represented with an ordered list of nodes~\cite{OSMnode}. From the paths of the OSM, we know which nodes are adjacent, and once a node is reached, we know which nodes we can move towards next. For example, a four-way intersection will have a single node at the intersection and four adjacent nodes. Since we reach the intersection from one direction, we have three possible next nodes from which to choose. The distance between road segments and the angles required to make a turn from the latitude and longitude pairs of the nodes can also be calculated.

OSM also includes other information such as speed limits that are useful for path prediction. Having speed limits for paths should improve the algorithm accuracy and efficiency by eliminating impossible candidate paths given speed trace. Other information included in OSM such as turn restrictions and way types are also helpful for the algorithm.

At the same time, time-stamped speed data also contain essential information. For example, we can easily obtain the driving distance and acceleration for the speed data. Therefore, a generic algorithm should use these quantities as part of the match criteria when fitting a speed trace to paths in the map.

\textbf{Adapting to Driving Restrictions:}
There are physical limitations to a vehicle's turning radius at a high speed. When a vehicle travels a particular path, it must travel at a speed at or below the maximum speed possible to make the turn. Also, when driving in a road, drivers need to obey speed limits. Although driving under maximum speed limit may not be always true for each driver, it still provides useful boundary cases for the algorithm.

During normal driving, people only stop when they have to. There are only a few scenarios when people would commonly stop: (1) they encounter red traffic lights or stop signs, (2) they start off from a place, (3) they reach to a destination, and (4) other cases such as road constructions, traffic, and pedestrians crossing the street. Our algorithm does not consider the last case since it happens randomly and it is not a common artifact in all traces.

\textbf{Using an Error Metric for Paths:}
A comparison metric is needed for the algorithm to determine the best path. There should be a scoring system representing how well a path matches with the speed trace.
It is difficult to find a path that perfectly matches the speed data, because drivers may swerve around objects in the road, or take turns more widely or sharply than we expect. Thus, the actual driving distance may not perfectly match with the path length calculated using the map. The challenge is that in some scenarios, these errors might cause even correct path to seem impossible. For example, if the model progresses past the 
vehicle's actual position along a segment of road, then the model may conclude that the vehicle is moving too quickly to make a
turn. If we corrected the distance traveled to account for some of these distance
estimation errors, then we may find that the speed traces line up perfectly with the turn. Thus,
any algorithm must correct paths while it explores them in an attempt to take into account these variations in the travel distance.

\subsection{Elastic Pathing Algorithm}
The elastic pathing algorithm is based on compressing or stretching the estimated distance that was traveled as we attempt to match the speed data to the path. We match zero speeds to intersections in the map and slow speeds to turns. Based on the daily driving scenario, these can only be one-way matching -- if the speed is zero, there needs to be an intersection; but if there is an intersection, the speed does not have to be zero (e.g.~when the traffic light is green). Similarly, if there is a turn, the speed needs to be slow enough to make the turn; but if the speed is slow (greater than zero), the car may not be turning. The compressing or stretching is done when there is a mismatch between the calculated distance from speed data and the length of a path segment in the map. Compressing means subtracting from the estimated distance by a specific amount. Stretching means adding to the estimated distance by a specific amount.

After reconciling differences between a section of road and the
speed trace, we must \textit{pin} the path at that point (which we call a landmark) because any movement 
would cause a mismatch between the two. For instance, if the driving speed in a trace is reduced to zero, indicating a
stop, where there is no intersection we might pull the path forward by some distance to reach an
intersection. All points that are pinned cannot be moved since
they align with features in the road. We call this approach \textit{elastic pathing} because the
stretching and compressing of the speed traces to fit the road is conceptually similar to stretching
a piece of elastic along a path while pinning it into place at different points. To understand the algorithm, we first introduce a set of definitions:

\begin{itemize}
\item \textbf{Calculated Distance:} The distance a vehicle traveled calculated from speed and
time values in the speed trace.
\item \textbf{Predicted Distance:} The distance along a possible route on the road at a certain time
in the speed trace.
\item \textbf{Error:} The difference between calculated distance and predicted distance of a possible
route.
\item \textbf{Feature:} A vehicle stop in the speed trace, an intersection in the road, or speed slow enough to make a turn.
\item \textbf{Landmark:} A place (e.g.~intersection or turn) where the speed trace and road data need to be checked to match.
\end{itemize}

\normalem
\begin{algorithm}[t!]
  \footnotesize
  \SetAlgoLined
  \DontPrintSemicolon
  \SetKwData {Samples}{Samples}
  \SetKwData {StartNode}{StartNode}
  \SetKwData {PartialPaths}{Partial}
  \SetKwData {CompletePaths}{Complete}
  \SetKwData {NewPaths}{P'}
  \SetKwData {TypeA}{TypeA}
  \SetKwData {ErrorThreshold}{$\delta$}
  \SetKwData {CurPath}{CurPath}
  \SetKwFunction{sort}{sort}
  \SetKwFunction{join}{join}
  \SetKwFunction{dropLowerQuantile}{dropLowerQuantile}
  \SetKwFunction{gotoBranch}{gotoBranch}
  \KwIn{The starting point-\StartNode, a set of speed samples- \Samples, and a
threshold within the best possible score to accept- \ErrorThreshold}
  \KwResult {\CompletePaths = list of possible paths within \ErrorThreshold of the best possible}
  \Begin {
    \PartialPaths $\longleftarrow \{[$\StartNode$]\}$\;
    \CompletePaths $\longleftarrow \emptyset$\;
    \While{\CompletePaths $= \emptyset$ OR \\\PartialPaths.first.error $< \ErrorThreshold \times$ \CompletePaths.first.error}{
      \tcp{Pop out the partial path with smallest error to advance}
      \NewPaths $\longleftarrow $ \gotoBranch(\PartialPaths.pop)\;
      \tcp{New paths are found after gotoBranch, then check for completed paths}
      \join(\CompletePaths, $\lbrace x \in \NewPaths \mid x$ complete $\rbrace$)\;
      \join(\PartialPaths, $\lbrace x \in \NewPaths \mid x$ incomplete $\rbrace$)\;
      \tcp{Sort by error: the first partial path (Partial.first) has the smallest error}
      \sort(\PartialPaths)
    }
    \tcp{The first completed path found has the smallest error}
  }
  \caption{\label{alg:pathing}Pseudocode for the elastic pathing algorithm. Starting with the StartNode, partial paths are added and processed through \emph{gotoBranch} function. In each iteration, partial paths are sorted so that the one with smallest error gets prioritized in search.}
\end{algorithm}

The pseudocode for our elastic pathing algorithm is given in
Algorithm~\ref{alg:pathing}. We measure the fit of a path by the amount of stretching or compressing that needs to be done
for the speed data to match the path. The more stretching or compressing along a path, the worse
this path scores. In Algorithm~\ref{alg:pathing}, the \textit{error} attribute for each candidate path measures this quantity. The algorithm stops exploring when it finds a set of possible solution paths (stored in ``Complete'' array) and the smallest error from other left-over uncompleted paths (stored in ``Partial'' array) is much worse than the best path that we have already found. 

At each iteration of the algorithm we sort all of the partial paths by their current
error and then explore the path with the smallest error (see while loop in Algorithm~\ref{alg:pathing}). This path is advanced until it reaches a feature
that requires the pin operation. At this point there may be multiple ways to advance the path so
several new paths may be created. Each new path's error is adjusted to reflect the stretching or
compressing of the distance traveled from the last pinned landmark. The algorithm proceeds to
the next iteration and follows the path with the smallest error, and thus, the first path
to finish cannot be worse than any other path. The value of parameter $\delta$ (greater than one) determines the number of finished paths we can obtain. This allows us to observe the top n paths. However, its value does not affect the selection of the best path, since the first solution path found is always the best. We set $\delta$ to be 1.1 when only interested in finding the best path.

\normalem
\begin{algorithm}[t!]
  \footnotesize
  \SetAlgoLined
  \DontPrintSemicolon
  \SetKwData {P}{P}
  \SetKwData {S}{S}
  \SetKwData {i}{i}
  \SetKwData {Back}{Back}
  \SetKwData {Fore}{Fore}
  \SetKwData {Edge}{Edge}
  \SetKwData {Branches}{P'}
  \SetKwFunction{Pin}{Pin}
  \SetKwFunction{advanceA}{compressA}
  \SetKwFunction{rewindA}{stretchA}
  \SetKwFunction{advanceB}{compressB}
  \SetKwFunction{rewindB}{stretchB}
  \SetKwFunction{join}{join}
  \SetKwFunction{intersection}{intersection}
  \SetKwFunction{maxSpeed}{maxSpeed}
  \KwIn{A path-\P, speed samples-\S, and the current index into the speed samples-\i}
  \While{\i $<$ \S.length}{
    \tcp{Match turns at intersections to slower speeds.}
    \If{\P at intersection}{
      \Branches $\longleftarrow \emptyset$\;
      \ForEach{$\Edge \in $\intersection(\P)}{
        \If{$\S[\i]$.speed $\leq$ \maxSpeed(\P, \Edge)}{
          \Pin(\P)\;
          \join(\Branches, $\P\cup\Edge$)
        }
        \Else{
          \Fore $\longleftarrow$ \advanceA(\P)\;
          \Back $\longleftarrow$ \rewindA(\P)\;
          \join(\Branches, \Fore)\;
          \join(\Branches, \Back)\;
        }
      }
      \Return \Branches\;
    }
    \tcp{Match 0 speeds to intersections.}
    \If{$\S[\i]$.speed $\approx 0$}{
      \While{$\S[\i]$.speed $\approx 0$ AND\\\i $<$ \S.length}{
        $++\i$\;
      }
      \If{\P at intersection}{
        \Pin(\P)\;
        \Return \P\;
      }
      \Else{
        \Fore $\longleftarrow$ \advanceB(\P)\;
        \Back $\longleftarrow$ \rewindB(\P)\;
        \Return \{\Fore, \Back\}\;
      }
    }
    \tcp{Process normally}
    \If{speed higher than way max speed + 20}{
        drop the current path
    }
    \Else{
        update total traveled distance
    }
    $++\i$\;
  }
  \caption{\label{alg:gotobranch}Pseudocode for the \textit{gotoBranch} function used in the elastic pathing
algorithm. This code advances a single path until it reaches discrepancy between the speed trace and
road. When this happens the path is corrected with compression or stretching and the path is
\textit{pinned} at what we call a landmark and the path's error is recomputed.  Multiple possible paths
may be returned at intersections.}
\end{algorithm}

Our algorithm uses maximum speed limits from OSM to determine whether a candidate path is possible. We have attempted several methods to apply speed limits to our algorithm such as segmenting out steady speed intervals for speed limit checking, separating highway and non-highway segments, and segmenting transient speed intervals for way entrance speed checking. However, they did not work out well experimentally. In the end, a simple solution of comparing vehicular speed to the maximum speed limit worked well. If the speed exceeds the maximum limit by more than 20 mph, the path is no longer considered possible. After trying different upper bounds, we found that speed limit plus 20 mph provides the best result overall in our testing stage. This seemed to be the upper bound that drivers in our datasets were willing to exceed the speed limit on the roads traveled.

The most complicated operation done in the \textit{ElasticPathing} algorithm is the \textit{gotoBranch}
function, for which pseudocode appears in Algorithm~\ref{alg:gotobranch}. The \textit{gotoBranch} function first verifies that the current speed does not exceed the speed limit of the road by more than 20 mph. It then
advances the path until it comes to a feature on the map, finds every possible path branching from
that feature, and returns those new possible paths. There are two types of feature matching we consider: a zero speed for an intersection and a slow enough speed for a turn.

When a possible path calculated by the algorithm reaches an intersection it explores every possible direction. If a vehicle on the possible path is going at a speed that can move along the curve in the road then a landmark is set at that location with the
\textit{Pin} function. If the curve in the road is too great for the current speed, we can rewind the speed cursor by a few samples to find a past speed sample that is slow enough to make the turn. This is equivalent to compressing the traveled distance. The amount of compressed distance can be calculated and added to the path error. This is done through our \textit{compressA} function. We can also advance the speed cursor to find a speed sample that is slow enough to make the turn. Our \textit{stretchA} function is used in this situation.
Therefore, when there is a mismatch between speed samples and road data, there are two ways to resolve it, and thus, two new possible paths.

A similar situation occurs when the speed data indicates that the vehicle has come to a stop. If the
path is already at an intersection then the landmark is set with the \textit{Pin} function. Otherwise
the two solutions are found with the \textit{compressB} and \textit{stretchB} functions. These two functions do not require rewinding or advancing the speed cursor. They directly compare the calculated distance from speed data to the road distance in the map and compute the difference (or error). They are still based on the concept of compressing and stretching the distance but have different implementation compared with \textit{compressA} and \textit{stretchA} functions mentioned above.

After any landmarks are set, the amount of stretching or compressing from the last landmark increases
the error of each path. This means that an existing path may have less error than the current paths
and should be explored instead. Therefore, all possible branches are returned to the elastic pathing
function and the paths are placed in sorted order by their error. The pathing then continues with the new
best path.

There are several details that we do not show in the pseudo-code. For instance, when we check if we
are at an intersection, we allow space for the number of lanes in the road and
offset for another car in front of our vehicle. Determining the minimum safe turn radius for a speed value is done by
assuming the maximum allowed incline in the road (8\%) \cite{roesstrafficengineering}. The equation that we use comes from Roess et al.~\cite{roesstrafficengineering}, and it is given as:
{\small
\begin{equation}
r\_safe = \frac{speed^2 } {(15\times(0.01\times\mbox{DEFAULT ELEVATION}+\mbox{friction}))}
\label{eq:1}
\end{equation}
}
In the safe turn radius equation, friction is the dry coefficient of sliding side friction for tires. 
Maximum default elevation is assumed to be 8\% which is the maximum allowed for areas where ice can form on the road. Setting maximum elevation allows the minimum radius to make a safe turn, allowing more paths than rejecting them. 
The minimum safe turn radius for a speed is then compared with the actual turn radius of the road which can be determined based upon the number of lanes in the road and typical lane widths \cite{aasho}. If the actual turn radius is smaller than the safe turn radius, the vehicle cannot make the turn at the given speed.

\subsection{Other Implementation Details}
\label{subsec:AlgorithmEnhancements}
In this section, we discuss other key implementation details of our elastic pathing algorithm and improvements that we made over our initial work~\cite{Gao:2014:EPY:2632048.2632077}.

\begin{figure}[t!]
\includegraphics[width=7cm]{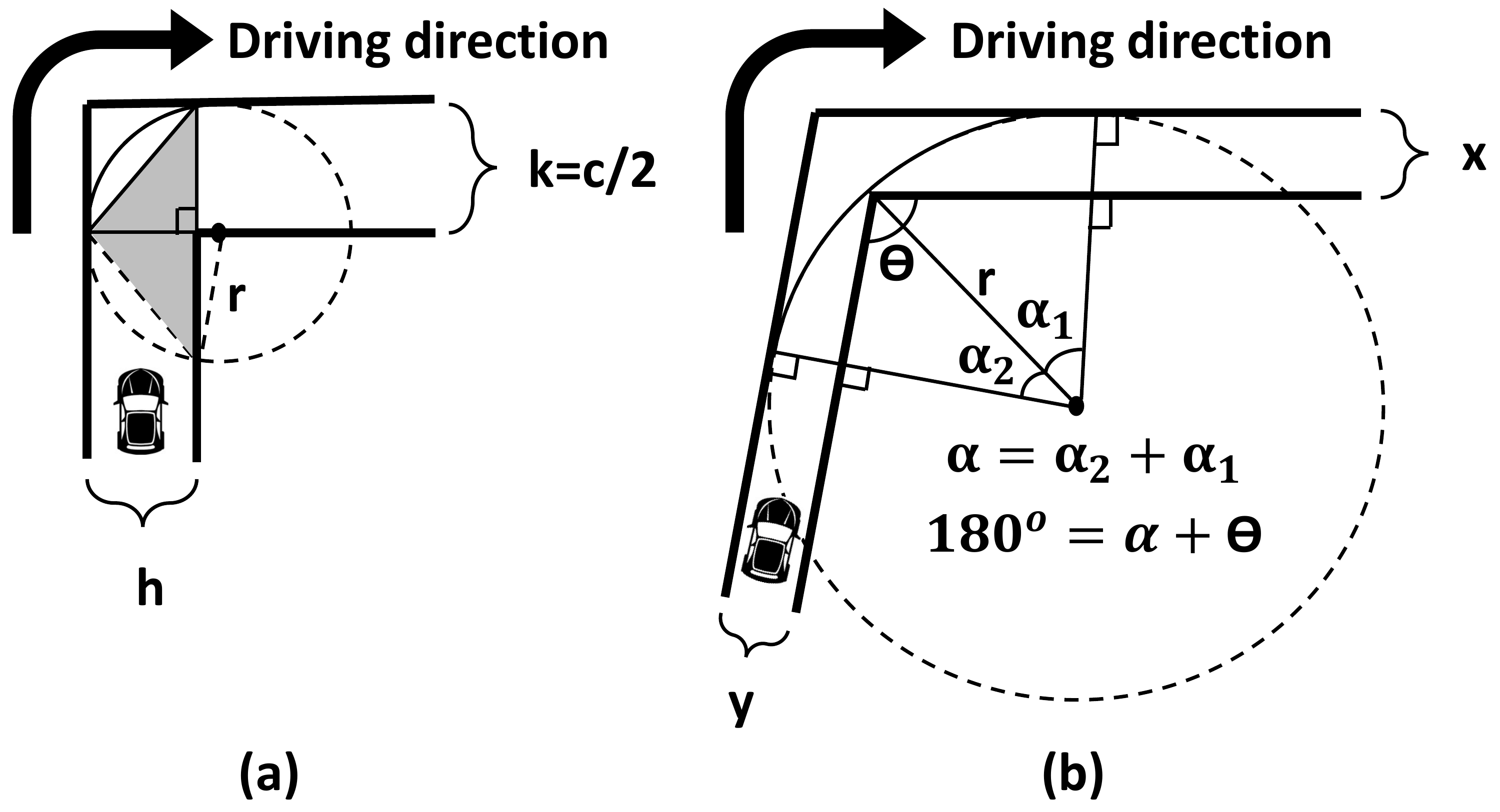}
\centering
\caption{\label{fig:turn_radius_fig} (a) The figure in the left presents our initial approach to estimate the turn radius which is set to be the radius of the circle inscribing the triangle (shaded gray in the figure). (b) The right figure shows our enhanced turn radius calculation method. The turning angle does not have to be 90 degrees and it has the maximum turn radius among all possible turns made within the path boundary.}
\end{figure}

\textbf{Applying a New Model for Turn Radius Determination:} The turn radius of the intersection determines the maximum driving speed at which a vehicle could make the turn. This provides an estimate of the feasibility that a driver can be in the turning location given the speed in the corresponding time instance. Thus a good model to determine the actual turn radius is essential to the algorithm.
Our previous attempt was to assume a sharp turn by drawing a circle to inscribe a triangle. Then, the radius of the circle was set to be the turn radius (see Figure~\ref{fig:turn_radius_fig}a): $r=h/2+c^2/(8h)$~\cite{Gao:2014:EPY:2632048.2632077}. Although a simple sharp turn happens frequently in most four-way intersections, we realized that a better model is necessary to account for different types of intersections (e.g.~three-way intersections, intersections with turning angles not equal to 90 degrees).
In addition, the resulting radius from the old model was not the maximum radius allowed given the geometric structure. This would cause the calculated maximum speed to be smaller than the actual possible speed, causing the algorithm to falsely reject candidate paths that may have been correct.

We used a new approach to model this turning event. Instead of having a sharp turn in the intersection, one can gradually turn from the outer side of the path, passing the intersection corner, and arriving to the outer edge of the other path (see Figure~\ref{fig:turn_radius_fig}b). This model gives the largest possible turning radius. In this model, the turning angle $\alpha$ does not have to be 90 degrees. The beginning path width $x$ and the ending path width $y$ can also be different. To solve for the turn radius $r$, we just need to apply some simple trigonometric functions and obtain the following equation:
{\small
\begin{equation}
    \cos{\alpha}=\frac{r-x}{r}\cdot\frac{r-y}{r}-\frac{\sqrt{r^2-(r-x)^2}}{r}\cdot\frac{\sqrt{r^2-(r-y)^2}}{r}
\end{equation}
}
Thus the turn radius $r$ can be solved (e.g.~there are two solutions, but we need the solution with larger value in this case), which is:
{\small
\begin{equation}
    r=\frac{x+y+\sqrt{2xy(1+\cos{\alpha})}}{1-\cos{\alpha}}
\end{equation}
}
This turn radius is then compared with the minimum safe turn radius 
to determine whether the speed is too fast to make the turn.

\textbf{Routing Method as an Optional Customization:} OpenStreetMap~\cite{routing_OSM} provides a routing (or navigating) feature to help users moving from one place to another.
The ruby version of the OSM routing is implemented in Mormon~\cite{mormon1} based on pyroutelib2~\cite{pyroutelib_ref}. OSM routing, based on A* search algorithm with weights, finds the shortest path between the two locations~\cite{mormon1}. 
We applied OSM routing as it is open source and provides free unlimited amount of usage on a given amount of time period.

The idea is to run our main elastic pathing search function first and use routing information to refine the score of our top candidate paths. After the main search function, there is a list of candidate paths ranked by their error.
We only consider the top 10 candidate paths to guarantee the quality of the selected candidate paths and avoid run-time overhead. Each candidate path has an estimated route, estimated destination, and an error. By comparing the length of a candidate path to the length of a navigated path, we can further evaluate the candidate path: the smaller the difference between lengths, the better the candidate path is.

This is under the assumption that people usually take an efficient driving route which should have a good match with the navigated route when comparing trip distance.
However, this may not be always true. For example, one may go with a much longer route to avoid heavy traffic. 
The effect on estimation accuracy of adding this method can only be tested with real-world datasets (see Section 5.5 for analysis).

We constructed our routing method as an optional function that could be easily switched on or off. 
To explain how we modified the error metric to consider the routing distance, we introduce the following parameters:
\begin{itemize}
\item \textbf{CalcDist:} This is the distance calculated from the speed and time values in the speed trace as defined previously. 
\item \textbf{RoutingDist:} This is the routing distance, shortest distance in this case, obtained using OSM routing method. 
\item \textbf{Error:} As defined previously, this is the difference between calculated distance and predicted distance, accumulated while compressing and expanding the candidate path. 
\item \textbf{MaxError:} This is the maximum \textbf{Error} among top 10 candidate paths for a driving trace.
\item \textbf{CombScore:} This is the final combined score used to determine how good a candidate path is -- larger score (ranging from 0 to 1) indicates better path.
\end{itemize}
To obtain the final combined score, we need to properly weight the error from the two factors: the difference between calculated distance and predicted distance, and the difference between calculated distance and routing distance.

There are three cases based on the value of the routing distance. 
The first case happens when the routing distance is smaller than the calculated distance. Since the routing distance is based on finding a shortest path, this case is expected to happen for some traces.
We apply the following formula to calculate the combined score:
{\small
\begin{equation}
\label{eq:6}
    CombScore=\beta\cdot\frac{MaxError-Error}{MaxError}+\omega\cdot\frac{RoutingDist}{CalcDist}
\end{equation}
}
where $\beta$ and $\omega$ are weight coefficients for these two fractions such that $\beta\in[0,1]$, $\omega\in[0,1]$, and $\beta+\omega=1$. The first fraction is simply the same as $1-(Error/MaxError)$. $Error/MaxError$ gives how bad the trace error is in range of $[0,1]$ (the larger the value, the worse the trace). As we want the fraction to reflect how good a trace is (the larger the value, the better the trace), we use $1-(Error/MaxError)$ to convert to what we need. The second fraction we use for routing score is $RoutingDist/CalDist$. As $RoutingDist$ is smaller than $CalDist$, this gives value ranging in [0,1] with larger score representing better routing match. 

The second case is when routing distance is larger or equal to the calculated distance. Theoretically, the routing distance should not be larger than calculated distance as routing is the shortest one. This case may still happen considering the calculation rounding error and the possibility of map data not exactly matching the actual road length.
In this case, we use the following formula:
{\small
\begin{equation}
\label{eq:7}
    CombScore=\beta\cdot\frac{MaxError-Error}{MaxError}+\omega\cdot Ratio_2
\end{equation}
}
where
{\small
\begin{equation}
\label{eq:8}
    Ratio_2=1-\frac{RoutingDist-CalcDist}{CalcDist}
\end{equation}
}
$(RoutingDist-CalcDist)/CalcDist$ measures the proportion of the difference to the calculated trip distance: $CalcDist$. We restricted this range to also be in [0,1] (the larger the value, the worse the trace). Note that if the RoutingDist is more than twice as long as CalcDist, we set it to equal to twice of CalcDist as this is the worst case. Then, we similarly use $1-(RoutingDist-CalcDist)/CalcDist$ to convert to what we need: the larger the score, the better the trace.

The reason we use this formula ($Ratio_2$ formula) is to ensure we rate both conditions (routing distance being larger and shorter) in the same way. For example, if the $CalcDist$ is equal to 1 mile, we need to make sure the rating is the same when $RoutingDist$ is equal to 0.8 miles and when $RoutingDist$ is equal to 1.2 miles, as they are both 0.2 miles difference to the $CalcDist$. Inevitably, this requires cutting off the $RoutingDist$ at $2*CalcDist$. An alternative approach would be simply using $CalcDist/RoutingDist$ instead of our $Ratio_2$ formula, but such metric will have distorted rating and lose the consistency between these two conditions. 

The third case is when routing distance cannot be found. This case happens when the OSM fails to find any route between two given points. However, this may not mean that there is no route connecting two points based on our testing. Sometimes, OSM API just failed to find the routes even for regular nodes. Thus we leave out the routing factor from our formula in this case:
{\small
\begin{equation}
\label{eq:9}
    CombScore=\frac{MaxError-Error}{MaxError}
\end{equation}
}

In general, this combined score metric works for different types of routing methods (e.g.~shortest path, fastest path, path with light traffic), as it considers the matching with two distances in general. In this paper, we only use it for the OSM routing of shortest path due to the limitation of the OSM routing API.

\textbf{Other Enhancements:} We re-implemented the landmark setting function to match with the new model for turn radius. We also adjusted the way we duplicate landmarks for newly generated candidate paths. We did minor modifications on the error accumulation functions to rematch beginning and ending points while calculating error based on speed traces. Finally, we corrected some minor issues while moving the time index in the speed trace to eliminate any offset it may produce during the path expanding and compressing.

\textbf{Space and Time Complexity:} Our elastic pathing algorithm is essentially a search algorithm optimized with a priority search queue. Different from depth-first search and breadth-first search, our algorithm prioritize the search on paths that are most likely to be the solution path. Therefore, our algorithm does not require searching through the whole path space. The first path found is the best path, and the first n path found are the top n ranked paths.

To estimate the space and time complexity, we can assume that there are N paths within the range that can be reached by the given driving distance. There are K nodes for each path (recall that nodes are locations representing a path). There are M samples in the given speed trace. M is much larger than K. The worst case happens when all paths are equally bad matches. The algorithm has to advance and shift among all the paths, resulting exploring all the paths within the range. The time complexity is M$\times$N, which can be noted as O(MN). The algorithm needs to at least store all the paths and the speed trace, so the space complexity in this case needs to be M+K$\times$N. The best case happens when there is only one distinct path matching well with the given speed trace. Then, the time complexity is about M, or O(M). The space complexity is about M+K. The average case time and space complexity of our algorithm depends on the actual driving environment or the path structure. With our datasets, fewer than 100 paths are explored in average based on the examination of our candidate path array list.

\subsection{Visualization Tool}
We built a visualization tool to show how our algorithm matches a speed trace to different paths along the map. The visualization interface was implemented using HTML5 and JavaScript for its convenience to host either locally or on the web. It takes an input file generated from our elastic pathing algorithm recording how each decision was made on each step. Then, using Google Maps APIs, we transferred the input data into the map with animation showing how our algorithm explores and selects paths.
The ground truth GPS data were plotted in the map with markers showing the starting location and the current position where the algorithm was exploring. Several top candidate paths that were explored upto the current time were also shown in the map.

The visualization of our algorithm is essential for the algorithm testing and debugging. Since our visualization tool takes separate input files instead of tying the implementation to our algorithm, it can be easily customized to similar applications that require the display of driving routes in the map.
\section{Driving Data}
To test how well our algorithm can reconstruct a driving route from a known starting point and a speed trace, we applied driving traces with different drivers and various driving environments. We required both a ground truth of GPS data and a sample set of speed data. This data was collected with the approach that the insurance companies use, by connecting a device to the On-Board Diagnostics standard (OBD-II)
connector. To log the required data we used two devices: a GPS-enabled smartphone, and a Bluetooth-enabled ODB-II device. We recorded the timestamped speed data and the corresponding GPS positions simultaneously. We also obtained a much higher volume of GPS-only data, from which we reconstructed the speed traces.

To determine the appropriate sampling rate for data collection, we referred to how insurance companies collect speed data. Although companies do not directly disclose their data sampling rate on their websites,
 Allstate~ \cite{allstatedrivewise} stated that ``hard braking events are recorded when your vehicle decelerates
more than 8 mph in one second (11.7 $ft/s^2$); extreme braking events are recorded when your
vehicle decelerates more than 10 mph in one second (14.6 $ft/s^2$).''  
We know that the sampling rate must be faster than the duration of the 
feature the insurance company wishes to observe. In the United States, federal vehicular safety 
regulations mandate that vehicles traveling at 20 mph must be able to
brake to a full stop in 20 feet at a deceleration rate of 21 feet/second$^2$~\cite{USmotorsafety393}.
These events that should be detected occur over a time interval of one second ($\tau = 1$). Using the Nyquist 
sampling theorem, we know that they must sample at twice this rate (two samples per second) to detect these features. To ensure the lower bound of sampling rates any company may use, we used sampling rate of one sample per second. This is much slower than what insurance companies use. Slower sampling rate means less information due to fewer samples, thus it is harder to estimate routes.

Using the speedometer data from the OBD-II device was straightforward because raw speed data was
immediately available. 
However, some driving traces were only available as GPS traces. This data required additional processing to obtain speed values
from latitude and longitude pairs. This was a two step process: (1) we applied the haversine formula~\cite{sinnott1984sky} to approximate distances from the raw GPS coordinates;
(2) we divided this value by the time interval to get an instantaneous speed value for each time interval. Each speed value was then given a timestamp and was converted into the same format as the speedometer traces. We note that we did not see differences in the accuracy of our algorithm between the two approaches of data collection. We also verified that the GPS data approximates actual speedometer data very well, and thus, does not affect our results.

\begin{figure}[t!]
\includegraphics[width=7.5cm]{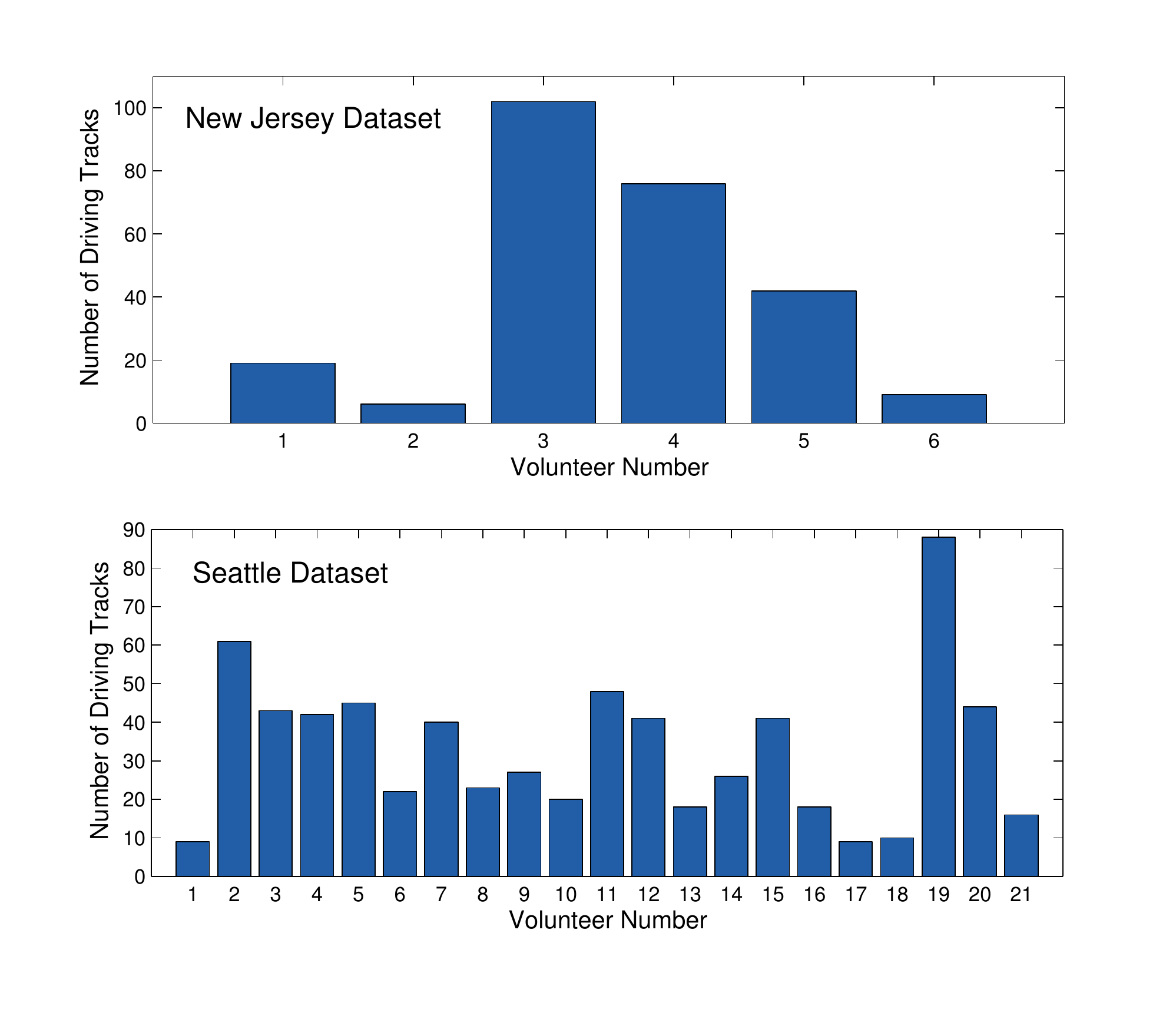}
\centering
\caption{\label{fig:Seattle_Individual1} The number of driving traces for each volunteer in the New Jersey dataset (top figure) and the Seattle dataset (bottom figure).
}
\end{figure}

After obtaining driving data, we slip them into driving traces as units of inputs to our algorithm. Our algorithm assumes that each driving trace does not have intermediate stops, meaning that all stops between a start location and a destination can only be caused by road conditions (e.g.~traffic light, stop sign, or turns).
We split the trip when the driver location does not change for more than 5 minutes, 
since stops due to traffic conditions are usually much less than 5 minutes.

We used three datasets to evaluate how our algorithm works with the real-world driving traces: two datasets, the central New Jersey dataset representing suburban areas and the Seattle city dataset representing urban areas with nearly one thousand testing traces in total, to estimate the overall estimation accuracy of our algorithm, and one identical-route dataset to further explore how different driving behaviors may affect our algorithm performance. The central New Jersey dataset and the Seattle dataset were used to test our algorithm in our initial work~\cite{Gao:2014:EPY:2632048.2632077}.  The third data set, named the identical-route dataset, was newly added in this paper.

\subsection{Central New Jersey Dataset}
We had six volunteers collecting data. Four volunteers collected GPS-only data over a period of three months, and two volunteers gathered both GPS and speedometer data with an OBD-II device over a period of one month. Examples of vehicles that were used in New Jersey dataset collection were small sedans, sports utility vehicles, and a pickup truck. The sampling rate for both the GPS and the speedometer was one second. Figure~\ref{fig:Seattle_Individual1} shows counts of individual driver traces collected. There were 254 data traces that we used for pathing with nearly 1250 miles (nearly 2012 km) in total, with a median trip length of 4.65 miles (7.5 km), minimum trip distance of 0.38 miles (0.62 km), and maximum trip length of 9.96 miles (16.0 km). These traces comprise 46 unique destinations, with more than half of these destinations visited multiple times by individual drivers. 28 locations were visited more than once during data collection and ten locations were
visited more than five times. The total driving time for 254 data traces was about 77 hours with an average driving time of 18 minutes for average driving distance of 4.9 miles.
The driving areas were mostly not dense urban areas, but residential, suburban and commercial areas connected by highways.

\subsection{Seattle Dataset}
To test our algorithm performance on denser urban areas, we used an external GPS dataset from Microsoft Research (MSR)~\cite{MSR_GPS}. The MSR GPS data was collected by 21 volunteers carrying a GPS logger for about eight weeks in the fall of 2009 in the region of Seattle, Washington. We had looked through several datasets available for access from various research centers and selected this particular dataset for its large size and great potential of extracting valuable driving traces for our testing. However, not all the traces were driving traces, since volunteers also carried around the GPS logger while they were walking or traveling by train or plane. This required the design of a trace filter to pick out all the driving traces. We extracted all the GPS traces that matched the following criteria: the trace consisted of driving in Seattle, and lasted for at least three minutes. This was done in order to have a reasonable distance with all traces and exclude trivial examples.

Another issue we needed to consider with this dataset was how to handle the loss of GPS signal while driving. For example, a car driving in a tunnel may lose GPS signal temporally. To ensure the accuracy of the speed calculation using GPS trajectories,  two sequential GPS readings should not be separated by more than five seconds. If this happens, they should be considered as two different driving traces. The possibility of losing signal is the only disadvantage of using a GPS device instead of using a speedometer. The dataset from New Jersey did not have this issue. 

After trace filtering, 691 traces were extracted with total driving distance of 1778 miles. Figure~\ref{fig:Seattle_Individual1} shows the counts of traces per participant. The mean driving distance per trace was 2.6 miles with minimum 0.59 miles and maximum 19.9 miles. 
The average driving time per trace was 11 minutes. Comparing with our own dataset collected by six volunteers, the Seattle dataset has relatively shorter average driving distance. Since it is an urban area, the participants' average driving distance per trip may be shorter than in suburban areas. We had no control of or information on the vehicles used since the dataset was from MSR and was collected independently to our research efforts. 

\subsection{New Identical-Route Dataset}
To investigate how individual difference may affect our algorithm accuracy, we recruited 30 participants (18 males and 12 females) to collect driving data. 20 of them were recruited in December 2014 and 10 were recruited in January and February of 2015. All of them were at least 18 years old and had a valid driving license. They were required to be an active driver during the past three months. 
Our participants are were all undergraduate and graduate students. 
Each of them was compensated with a \$50 visa gift card for completing the driving study. We labeled these participants as P1, P2, ..., P30.

\begin{figure}[!t]
\centering
\includegraphics[height=5.5cm]{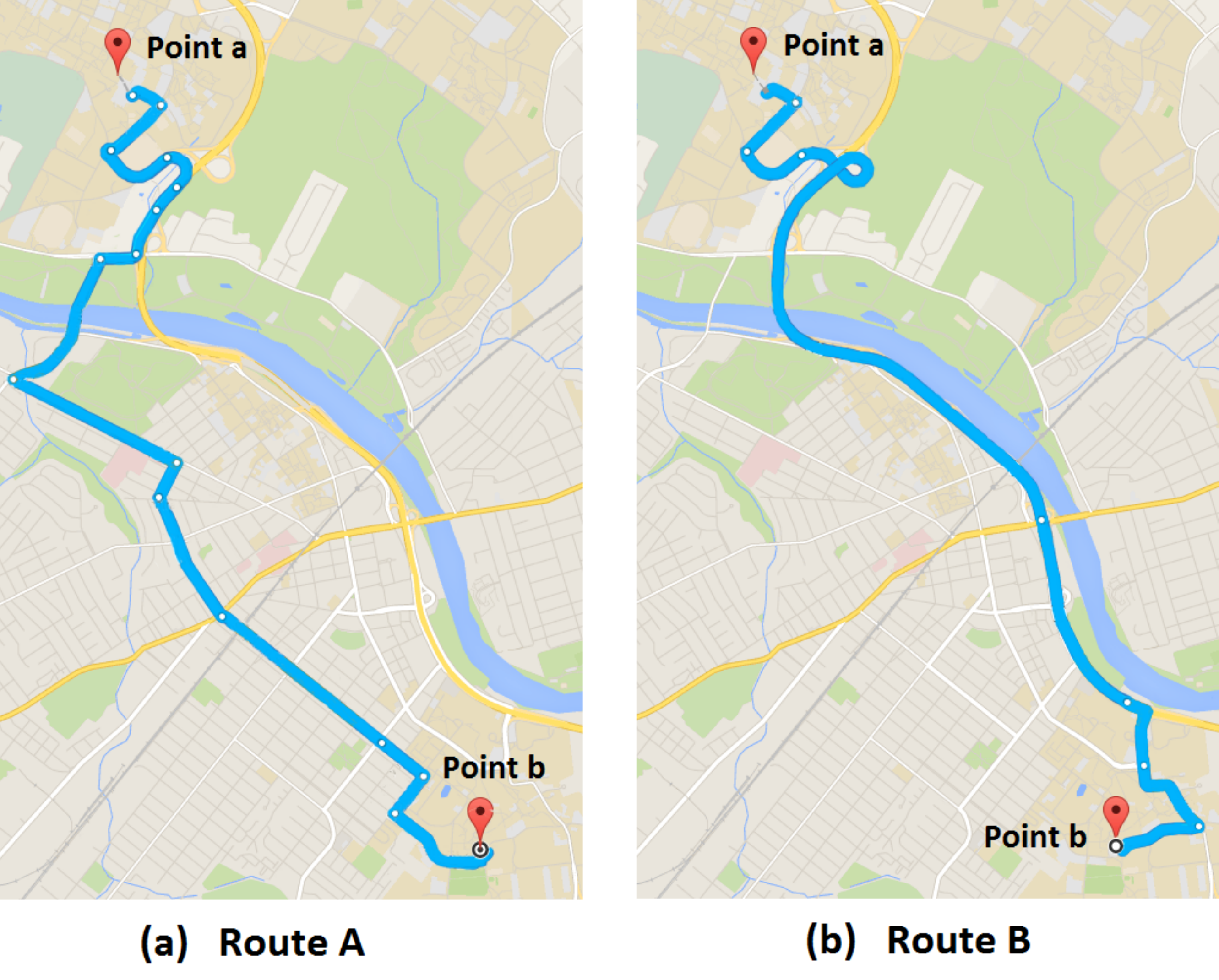}\caption{Route A shows the driving route from point a to point b, which comprises residential and commercial areas with some traffic lights. Route B is from point b back to point a, and it consists of a long highway segment which connects local paths with a few traffic lights.}
\label{fig:route}
\end{figure}

We required all participants to drive in the same route twice. The route was for a round trip where participants needed to drive from point a to point b, and then return to point a from a different path. We split this round-trip route into the leaving route (denoted as route A) and the returning route (denoted as route B). Thus both route A and route B were traveled twice for each participant. These two routes (see Figure~\ref{fig:route}) were across different areas and have different driving environments: Route A consists of more traffic lights and intersections than route B.

A 2002 Hyundai Accent car was used for all the driving traces in this dataset. Both the speed data and GPS trace were collected using a smartphone and an OBD-II connector. Before data collection, participants were allowed to drive with the test car in the corresponding area to get familiar with the vehicle and the driving environment. This was done to eliminate any adjustment or learning effects from consideration.
To avoid heavy traffic and ensure similar driving condition, all the driving traces were collected during late morning and early afternoon with similar weather condition (e.g.~either slightly cloudy or sunny). 

In this dataset we used a sampling rate of one sample per second. There are 60 driving traces in route A (4.8 miles) and 60 driving traces in route B (5.1 miles) by 30 different drivers. However, one participant (e.g.~P19) once drove to a wrong route during the study, resulting only 59 valid driving traces for route A.

\section{Results}
We present results of our algorithm in this section. We start by presenting the overall accuracy of our algorithm when testing with two comprehensive datasets: New Jersey dataset (representing sub-urban areas) and Seattle dataset (representing urban areas). We discuss the effect of applying OSM routing and road speed limits by comparing estimation accuracy. We then investigate how different drivers and driving behaviors may affect our algorithm by analyzing the result from our new dataset: identical-route dataset. Finally, we analyze the traces with poor estimation accuracy and point out major factors that may mislead our algorithm.

\begin{table}[]
\centering
\small
\begin{tabular}{|p{1.4cm}|p{1.25cm}|p{0.8cm}|p{0.9cm}|p{0.8cm}|p{0.9cm}|}
\hline
\multicolumn{2}{|p{2.65cm}|}{Destination Error} & \multicolumn{2}{p{1.7cm}|}{New Jersey} & \multicolumn{2}{p{1.7cm}|}{Seattle}\\
\hline
Meters&Miles&Trace Count&Percent&Trace Count&Percent\\
\hline
0-250 & 0-0.16 & 44 & 17.32\% & 107 & 15.48\%\\
\hline
250-500 & 0.16-0.31 & 18 & 7.09\% & 83 & 12.01\%\\
\hline
500-750 & 0.31-0.47 & 15 & 5.91\% & 82 & 11.87\%\\
\hline
750-1000 & 0.47-0.62 & 12 & 4.72\% & 65 & 9.41\%\\
\hline
1000-1250 & 0.62-0.78 & 7 & 2.76\% & 43 & 6.22\%\\
\hline
1250-1500 & 0.78-0.93 & 14 & 5.51\% & 26 & 3.76\%\\
\hline
1500-1750 & 0.93-1.09 & 7 & 2.76\% & 31 & 4.49\%\\
\hline
1750-2000 & 1.09-1.24 & 14 & 5.51\% & 34 & 4.92\%\\
\hline
2000-2250 & 1.24-1.40 & 8 & 3.15\% & 17 & 2.46\%\\
\hline
2250-2500 & 1.40-1.55 & 10 & 3.94\% & 16 & 2.32\%\\
\hline
2500-2750 & 1.55-1.71 & 3 & 1.18\% & 26 & 3.76\%\\
\hline
2750-3000 & 1.71-1.86 & 4 & 1.57\% & 9 & 1.30\%\\
\hline
3000-3250 & 1.86-2.02 & 3 & 1.18\% & 10 & 1.45\%\\
\hline
\textgreater  3250 & \textgreater  2.02 & 95 & 37.40\% & 142 & 20.55\%\\
\hline
\multicolumn{2}{|p{2.65cm}|}{Total traces} & 254 & 100\% & 691 & 100\%\\
\hline
\end{tabular}
\vspace{1pt}
\caption{\label{table:new_results} Results from the definite version of our elastic pathing algorithm for both the New Jersey and the Seattle dataset. Table shows the number of traces having destination error ranging from 0 to greater than
3.25 km, separated into 14 intervals as shown in rows. First two columns give destination error representations in meters and miles.}
\end{table}

\subsection{Overall Accuracy}
The Ruby implementation of
our algorithm processed all of the New Jersey
traces (254 traces) in about one hour on a two-core 2.2
GHz machine. For the Seattle dataset (691 traces), it took
about two hours. With nearly one thousand traces in total, about 80\% of them had running time within just a couple of seconds per trace while remaining ones took much longer and usually had much worse accuracy. 
The processing speed for traces can be done much faster than the rate at which those same traces are generated. This may enable real-time tracking of drivers with just speed data, however it will not be 100\% accurate.

Roughly half of the traces can be estimated with destination error less than one mile for both the New Jersey dataset and the Seattle dataset (see Table~\ref{table:new_results}). A significant percent of traces can be estimated with very high accuracy: 17\% of traces with destination error less than 250 meters and 24\% of traces with destination error less than 500 meters for the New Jersey dataset, and 15.5\% and 27.5\% of traces with destination error less than 250 meters and 500 meters respectively for the Seattle dataset.
 There is a noticeable improvement (especially for error less than 250 meters) in the estimation accuracy compared to our initial version~\cite{Gao:2014:EPY:2632048.2632077}~(see Table~\ref{table:comparisonWithOldResult} for comparison). Overall, 150 traces having destination error within 250 meters, comparing to 125 traces in our initial work: there are 25 more traces (or 20 percent improvement of 125 traces) in this highly accurate estimation range.

\begin{table}[]
\centering
\small
\begin{tabular}{|m{1.4cm}|m{1.2cm}|m{1.2cm}|m{1.2cm}|m{1.2cm}|}
\hline
\multirow{2}{0.35in}{Destination Error (meters)} & \multicolumn{2}{m{2.4cm}|}{New Jersey} & \multicolumn{2}{m{2.4cm}|}{Seattle}\\\cline{2-5}
&Initial \newline Version & Definite \newline Version & Initial \newline Version & Definite \newline Version\\
\hline
\textless 250 & 14.17\% & 17.32\% & 13.02\% & 15.48\%\\
\hline
\textless 500 & 23.62\% & 24.41\% & 26.34\% & 27.50\%\\
\hline
\textless 750 & 29.13\% & 30.31\% & 34.88\% & 39.36\%\\
\hline
\textless 1000 & 33.46\% & 35.04\% & 43.42\% & 48.77\%\\
\hline
\textless 1250 & 37.80\% & 37.80\% & 50.51\% & 54.99\%\\
\hline
\end{tabular}
\caption{\label{table:comparisonWithOldResult} Comparison of estimation accuracy between two versions of our algorithm. Table shows percentages of traces that has destination error within certain range for both New Jersey and Seattle dataset.}
\end{table}

\begin{figure}
\centering
\includegraphics[width=7cm]{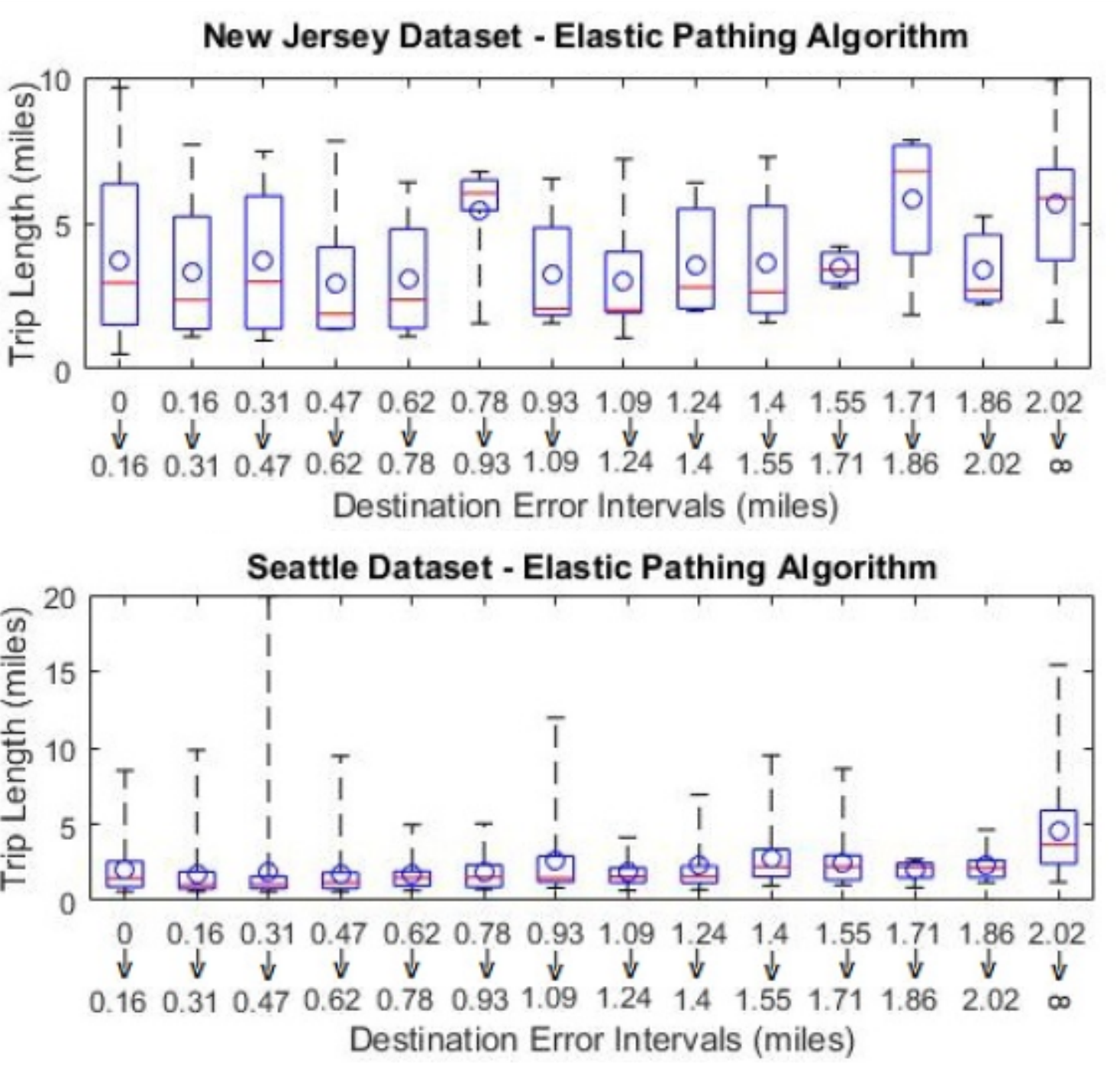}
\caption{Box plots of estimation accuracy (or destination error) vs. driving trip length for the definite version of elastic pathing algorithm. The destination error is split into 14 intervals (e.g.~interval 0-0.16 miles, interval 0.16-0.31 miles) with each interval equal 0.155 miles or (250 meters). Trip length distribution for each error interval is presented with a box-plot bar, showing: minimum, maximum, first quartile to third quartile range (blue box), median (red line), and mean (blue circle).}
\label{fig:trip_length_dest_error}
\end{figure}

To provide additional context of our results from definite version of our elastic pathing algorithm, we analyze how estimation accuracy depends on the trip length. Figure~\ref{fig:trip_length_dest_error} shows the distribution of trip lengths within each interval of estimation accuracy -- from highly accurate interval (destination error less than 0.16 miles or 250 meters) to very rough estimation interval (error greater than 2 miles). In New Jersey dataset, for traces estimated with high accuracy (error less than 0.16 mile or 250 meters), trip length varies from 0.38 miles to 9.64 miles. This means that not only can our algorithm predict short traces with good accuracy, it can also predict very long traces with good accuracy. The average trip length for those with error less than 0.16 mile is about 4 miles. While the average trip length varies for traces in different error intervals, we did not see any clear dependence of our algorithm's performance to the trip length based on the figure. For Seattle dataset, the general variance of trip length is smaller than the one in New Jersey dataset, since Seattle traces have much shorter average trip length. Similarly, our algorithm can predict traces with both short and long trip lengths with good accuracy. While the average trip length for error larger than 2 miles is slightly longer than others, we did not find any consistent trend of trip length being larger in larger error interval. For example, traces in 0-0.16 mile interval has longer average trip length than those in 1.71-1.86 mile interval; the longest trace with 20 mile length has error between 0.31 and 0.47 mile.

\subsection{Comparison with Naive Guessing}
We implemented a naive guessing algorithm to provide a baseline accuracy to compare with elastic pathing algorithm. The naive guessing algorithm takes the map data, a starting location, and the speed data as inputs, but it does not utilize any logic about matching zero speeds to intersections or any scoring metric.
From the starting location, the algorithm calculates the traveled distance from the speed data and selects a path randomly to advance. When encountering an intersection, the algorithm randomly selects a path to continue -- each direction has an equal chance of selection.

We found that our elastic pathing algorithm achieves much higher estimation accuracy than naive guessing algorithm. Since naive guessing algorithm is based upon random selection, the solution path can be different each time the algorithm runs. We ran the guessing algorithm for ten times to obtain a rough range of its estimation accuracy. For destination error within 250 meters, naive guessing algorithm only estimates 0.87\% (average value with min 0\% and max 2.36\%) of traces in New Jersey dataset and 4.55\% (average value with min 4.05\% and max 5.09\%) of traces in Seattle dataset. This is much lower than our elastic pathing algorithm's accuracy: 17.32\% for New Jersey dataset and 15.48\% for Seattle dataset. We can see that naive guessing algorithm has higher accuracy in Seattle dataset than New Jersey dataset. This is due to Seattle dataset has much shorter average trip length than New Jersey dataset (2.6 miles vs.~4.65 miles).

\subsection{Effect of Applying OSM Routing}
We applied the routing (or navigating) method into our algorithm and compared its estimation accuracy with the version without it. The routing feature is provided by the OSM API. In order to obtain the combined score for each driving trace, we need to determine the constants: $\beta$ which is the weight of the error accumulated by compressing and extending paths, and $\omega$ (equivalent to $1-\beta$) which is the weight of the routing error (see Equations~\ref{eq:6} to~\ref{eq:9} in Section~\ref{subsec:AlgorithmEnhancements}). We enumerated $\beta$ from 0 to 1 with 0.1 incremental steps. One extreme case is when $\beta$ equals 0. This means that only the routing error is considered to re-rank the top several candidate paths selected by our algorithm. Another extreme case is when $\beta$ equals 1. This is when OSM routing error is not used at all.

We found that applying OSM routing drops the overall accuracy for both New Jersey and Seattle datasets. Varying the value of $\beta$ from 0.1 to 0.9 does not have noticeable difference for the overall accuracy. Table~\ref{table:comparisonWithRouting} (``Add Routing'') shows the estimation accuracy when both $\beta$ and $\omega$ are set to 0.5. As shown in the table, our definite version without routing (happens when $\beta$=1) has better overall accuracy.

We found that OSM routing affects different drivers differently,  
especially for the New Jersey suburban dataset. There are six drivers in the New Jersey dataset. When $\beta$ changes from 0.5 to 1 for P1 and P3, the portion of driving traces having error within 500 meters increases (P1: 10\% to 20\%, P3: 18\% to 32\%), meaning that our algorithm has higher estimation accuracy without using OSM routing. However, for P4 and P6, OSM routing helps increase the estimation accuracy for error less than 500 meters (P4: 22\% to 28\%, P6: 21\% to 45\%). For Seattle dataset, the accuracy of driving traces from 19 out of 21 drivers do not vary with $\beta$ value. In other words, routing calculation mostly agrees with the best path selected by elastic pathing algorithm. From the reminding two drivers, OSM routing increases the estimation accuracy for P13 but decreases the accuracy for P15.

\begin{table}[]
\centering
\small
\begin{tabular}{|m{1.8cm}|m{1.1cm}|m{1.1cm}|m{1.1cm}|m{1.1cm}|}
\hline
$\beta$=0.5, $\omega$=0.5 & \multicolumn{2}{m{2.2cm}|}{New Jersey} & \multicolumn{2}{m{2.2cm}|}{Seattle}\\
\hline
Destination \newline Error (meters) & Definite Version & Add Routing & Definite Version & Add Routing\\
\hline
\textless 250 & 17.32\% & 14.57\% & 15.48\% & 15.34\%\\
\hline
\textless 500 & 24.41\% & 20.08\% & 27.50\% & 27.35\%\\
\hline
\textless 750 & 30.31\% & 28.74\% & 39.36\% & 39.22\%\\
\hline
\textless 1000 & 35.04\% & 33.07\% & 48.77\% & 48.63\%\\
\hline
\textless 1250 & 37.80\% & 36.22\% & 54.99\% & 54.85\%\\
\hline
\end{tabular}
\vspace{1pt}
\caption{\label{table:comparisonWithRouting} Comparison of estimation accuracy when OSM routing is applied to our algorithm. Table shows percentages of traces that has destination error within certain range for both New Jersey and Seattle dataset.}
\end{table}

Applying OSM routing method also increases the algorithm execution time given that a corresponding navigated path needs to be found for each of the top ranked candidate paths. Our algorithm takes about 11 hours to process all the traces in the New Jersey dataset and 25 hours for Seattle dataset.

\subsection{Effect of Applying Road Speed Limits}
Our algorithm applied road speed limits obtained from OSM to determine how well a candidate path matches with a give speed trace. In this section, we explore how the algorithm performs when we relax this condition. This provides additional insights on how much the speed limit affects the estimation accuracy for different driving environments: urban and sub-urban.

\begin{figure}[t]
\centering
\includegraphics[width=\linewidth]{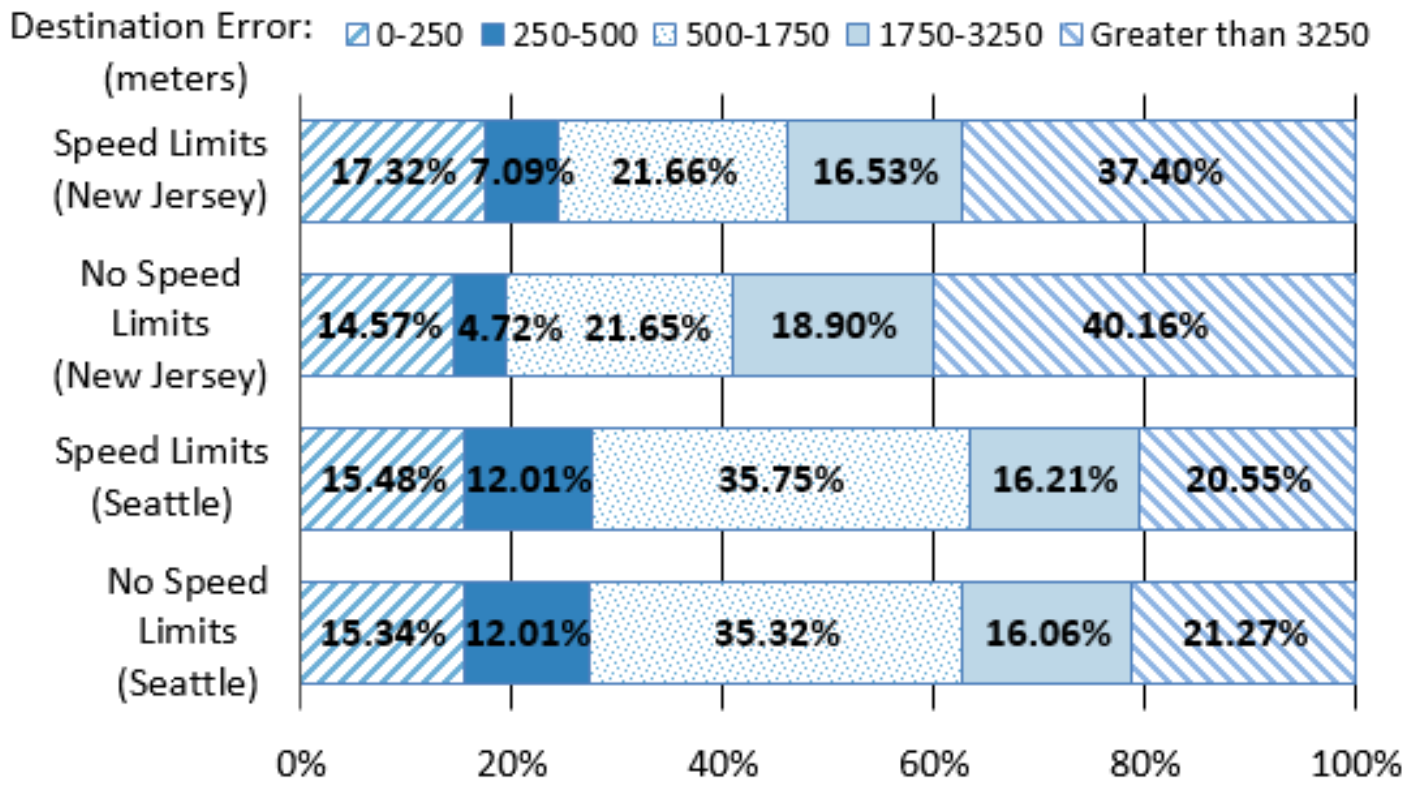}
\caption{Effect of applying speed limits on the New Jersey and the Seattle dataset. Figure shows the percent of driving traces from the corresponding dataset with destination error within different intervals: from 0-250 meters (0.16 mile) to greater than 3250 meters (2 miles).}
\label{fig:speed_limit_comparison}
\end{figure}

We found that without using speed limits, the estimation accuracy for Seattle dataset does not drop a lot (e.g.~only a few traces difference) while the accuracy drops noticeably by several percentage points for New Jersey dataset (see Figure~\ref{fig:speed_limit_comparison}). In New Jersey dataset, there are 14.57\% of traces (37 out of 254) having destination error within 250 meters when speed limits are not applied. This is 7 traces fewer than the one using speed limits. For error within 500 meters, there are 19.29\% of driving traces when speed limits are not applied (e.g.~13 traces fewer than the one with speed limits). Interestingly, for Seattle dataset, when we do not apply speed limits, the number of driving traces with destination error within 250 meters is only one trace fewer. The numbers of driving traces distributed in different error intervals are similar. 

\subsection{Results for Identical-Route Dataset}
As different drivers can have very different speed patterns even for the same route, we present results for the identical-route dataset with 30 drivers to investigate how driving behaviors may affect our algorithm.

\textbf{Overall Accuracy:} We processed all the driving traces (59 traces for route A and 60 traces for route B) in this dataset to see the distribution of destination errors. It took about 2 hours in total. We used the New Jersey map during the processing.

During the result analysis, we discovered and fixed some issues of the New Jersey map provided by OSM. (1) The nodes in route A and route B were not connected in a region (about 0.5 miles starting from point a in Figure~\ref{fig:route}). Thus, the routes were disconnected from the map, resulting the ground truth path never reached by our algorithm.
(2) There were other minor node connection issues: a short two-way road connecting to a local path in route A was mistakenly structured as one-way road in the map, a one-way road in route B was mistakenly structured as two-way road, and a node in one path was wrongly connected to a node in another path in route B.
(3) There were many rail ways connecting together with driving ways in the OSM, misleading the algorithm to explore rail ways as well.
The name of rail ways were sometimes not specified and sometimes not distinguishable from actual driving ways, making it difficult to separate with programming approaches. We fixed the nodes that were connected in route A and route B, making sure the ground truth path is reachable through the map. However, these issues may still happen in many regions of the map. 

\begin{figure}[t]
\centering
\includegraphics[width=\linewidth]{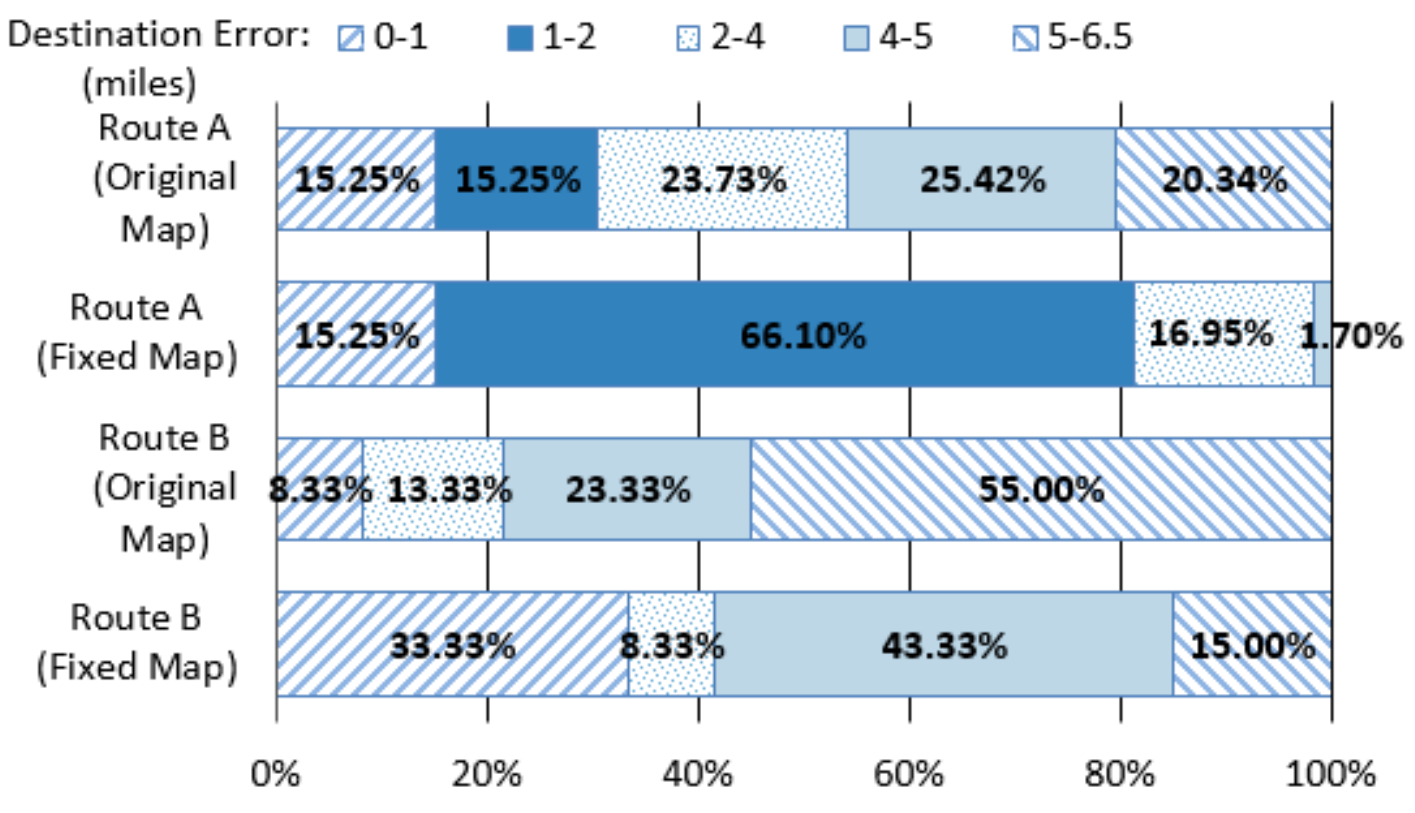}
\caption{Comparison of estimation accuracy when using the original map and the fixed map. Figure shows the percent of traces with destination error within different intervals: from 0-1 mile to 6.5 miles. There is no driving trace with destination error greater than 6.5 miles.}
\label{fig:identical-route-map-fixed}
\end{figure}

We compared the estimation accuracy of our algorithm using the original map and the fixed map (see Figure~\ref{fig:identical-route-map-fixed} for comparison). We found that fixing the map had a huge impact on the estimation accuracy for both route A and route B.
We did not see any difference on the overall accuracy when we applied this fixed map for the New Jersey dataset with 254 traces. This is not surprising, since we only fixed a small region related to the two routes. 

The overall estimation accuracy for the identical-route dataset is much lower than the New Jersey and Seattle datasets: 15\%-33\% traces with error less than 1 mile comparing to about 50\% traces with error less than 1 mile for New Jersey and Seattle datasets. This is because the identical-route dataset only consists of two driving routes with fixed region for all drivers. We selected these two driving routes to pass areas with driving environments that were shown to be difficult for our algorithm based on our prior work~\cite{Gao:2014:EPY:2632048.2632077}. These routes consists of highways, local and dense streets where there are many turns and intersections. We selected two relatively complex routes so that they can cover a large variety of driving environments and the results would have reasonable variation for different drivers to analyze how different drivers and different driving behaviors may affect our algorithm.

\textbf{Driving Behaviors vs. Estimation Accuracy:} Different driving behaviors should result in different patterns of speed traces. For example, drivers who drive faster would usually have higher average speed than others given the same route and similar traffic condition. Drivers who brake harder than others would have higher average braking deceleration. For both route A and route B, we extracted some common features from speed traces:
\begin{itemize}
\item \textbf{Average Speed:} This is the mean of speed values within one driving trace.
\item \textbf{Average Braking Deceleration:} We extracted all the monotonically decreasing speed intervals from a speed trace. For each speed interval with at least 10 mph decrement, we calculated the average deceleration. We further average these values from different intervals to obtain the overall average braking deceleration. Given only speed trace, it is not easy to extract only braking intervals without complicated analysis since the driving speed can also decrease when releasing the gas petal. Setting threshold for instant braking deceleration may bias some drivers. For our dataset, we found that 10 mph is a great threshold to filter out most non-breaking events for different drivers. 
\item \textbf{Number of Braking Events:} This is the number of monotonically decreasing speed intervals that have speed drop with at least 10 mph.
\item \textbf{Number of Stops:} This is the number of stops in one driving trace. We are including the starting and ending stops, so there are at least two stops for each trace.
\end{itemize}

\begin{figure}[t]
\centering
\includegraphics[width=8cm]{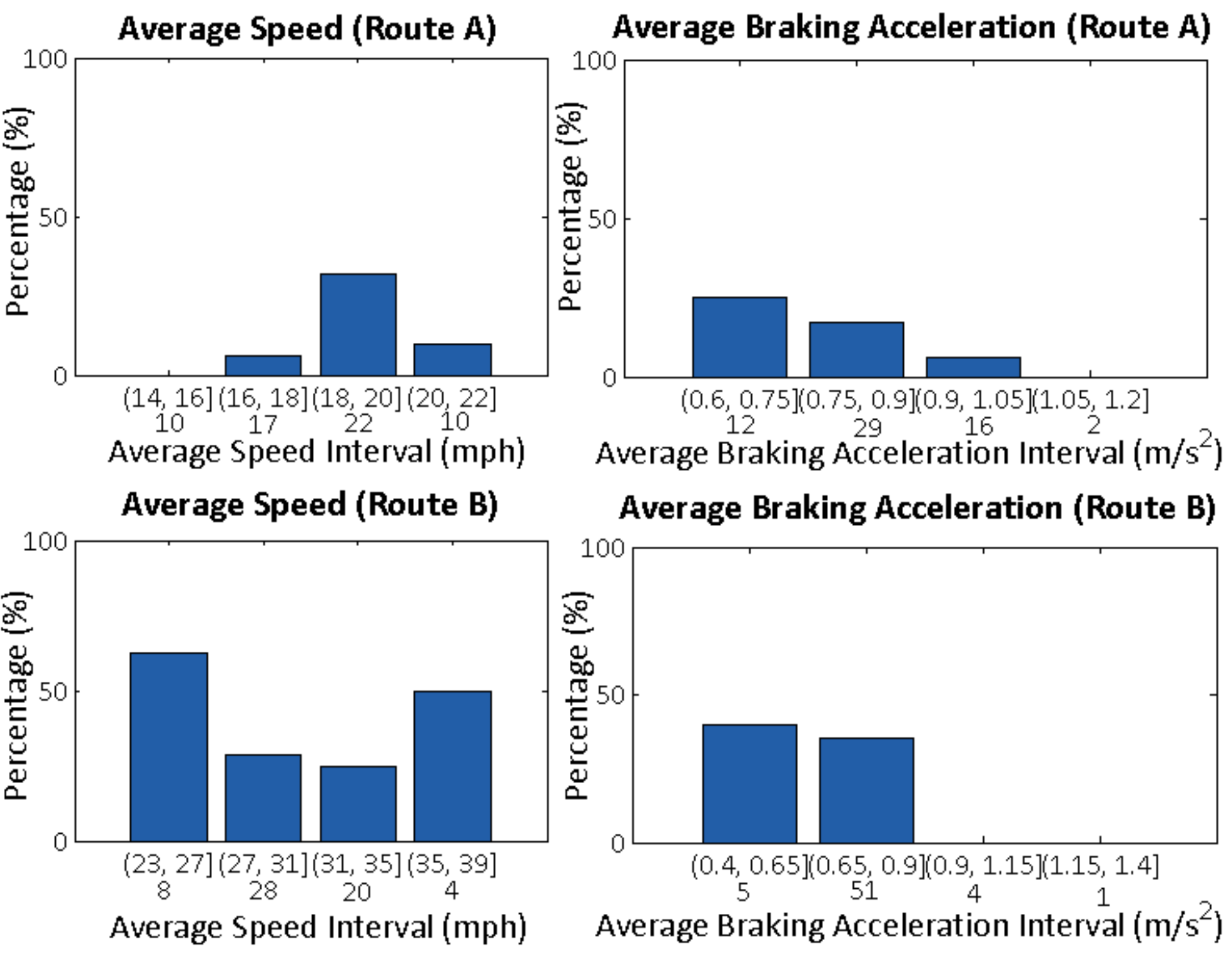}
\caption{The effect of average speed and average braking deceleration to the estimation accuracy. The range of average speed or average braking deceleration is split into four intervals. Values in horizontal axis show both the interval and the number of traces within the interval. The percentage value in vertical axis shows the percent of driving traces within each interval having destination error within one mile.}
\label{fig:identical-route1}
\end{figure}

Figure~\ref{fig:identical-route1} shows the percentage of driving traces that have destination error less than one mile for each interval of average speed and average braking deceleration. For driving traces corresponding to route A (e.g.~59 traces in total), there are 22 traces with average speed between 18 and 20 mph, and 31.8\% of them (7 out of 22) have destination error within one mile. This percentage is relatively high comparing to other speed intervals. Thus, our algorithm has slightly better estimation accuracy when the average speed is around 18 to 20 for route A. When the average speed is too slow or too fast, the probability of our algorithm providing good estimates drops rapidly. For instance, there is no trace with error within one mile among the 10 traces with average speed between 14 to 16 mph. For route B, the distribution of relatively accurate traces (e.g.~error less than one mile) over the average speed is different from route A's. The average speed is generally higher than route A. This is due to route B having a long segment of highway path while route A does not pass highways. Relatively low or high average speeds help on our algorithm's estimation in this case (e.g.~see figure~\ref{fig:identical-route1}). This result is also intuitive for route B: based on our trace analysis, slow average speed usually indicates more stops before or after the highway segment which then provides more information for our algorithm about intersections, and very fast average speed indicates high driving speed in highway segment which helps eliminate some unrelated local candidate paths during the algorithm's speed limit testing. When looking at average braking deceleration, our algorithm has higher probability of giving accurate estimation when the average deceleration is relatively low for both route A and route B.

\begin{figure}[t]
\centering
\includegraphics[width=8cm]{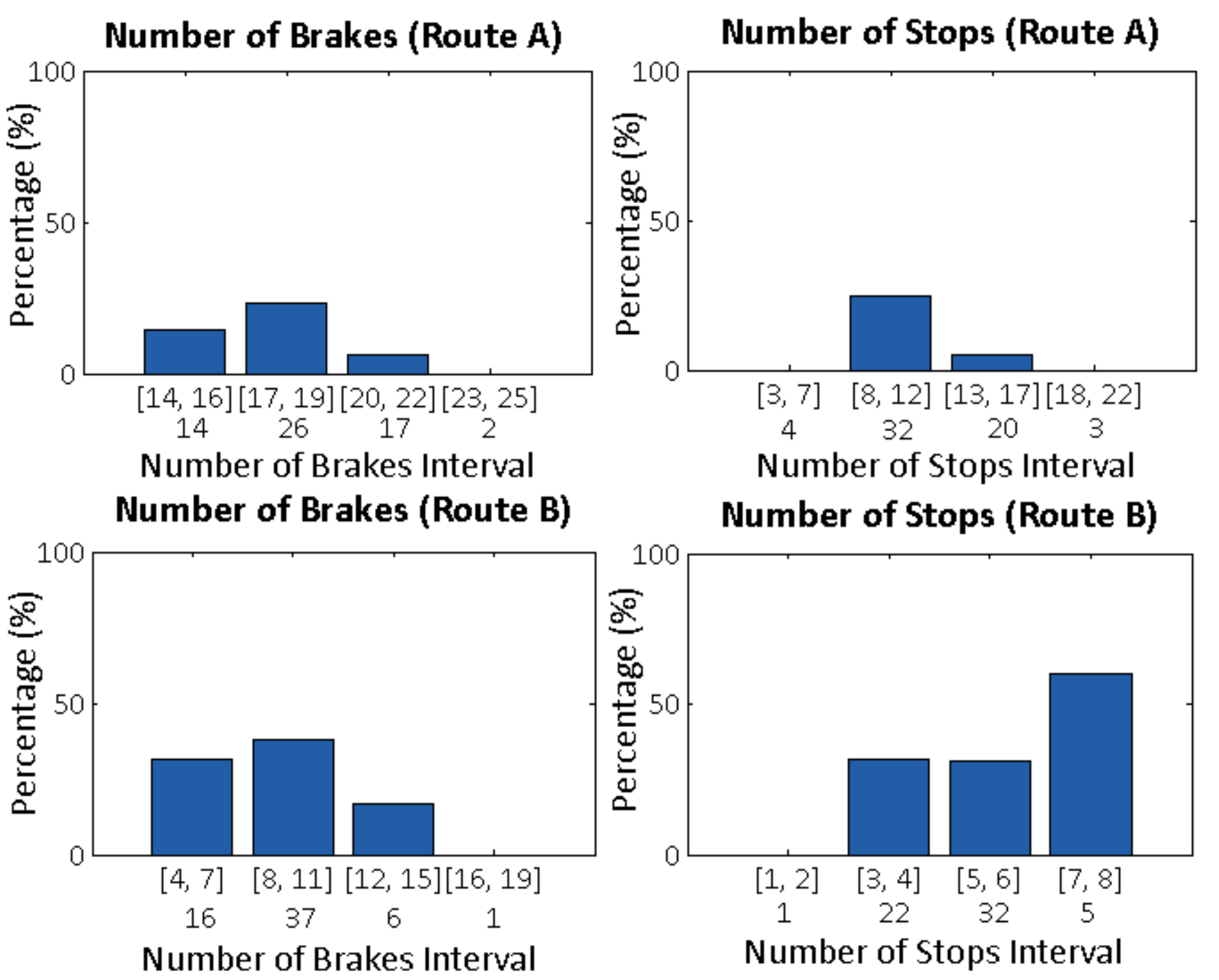}
\caption{The effect of brakes and stops to the estimation accuracy. We use ``brakes'' in the figure to denote ``braking events''. Each range is split into four intervals with integer values. Values in horizontal axis show both the interval and the number of traces within the interval. The percentage value in vertical axis shows the percent of traces within each interval having destination error within one mile.}
\label{fig:identical-route2}
\end{figure}

Figure~\ref{fig:identical-route2} shows how brakes and stops may affect algorithm's estimation accuracy. In general, drivers brake more often in route A than route B. For route A, majority of traces have a number of braking events between 17 to 19. These traces have the highest estimation accuracy. Similarly, for route B, the majority of traces have number of braking events between 8 to 11, and these traces also have the highest estimation accuracy. However, the distribution of accurate traces over the number of stops is different between route A and route B. 8 to 12 stops appear to be the best for our algorithm to estimate route A. While, more stops favors our algorithm's estimation in route B. We found that all 5 traces with 7 to 8 stops in route B have relatively low average speed (e.g.~23-27 mph in Figure~\ref{fig:identical-route1}). This aligns with the previous statement about why low average speed in route B may increase the estimation accuracy. We should, however, note that the data sets contained relatively few low and high average driving speeds, as these are extreme cases correspond to distinct driving styles.

We analyzed how brakes and stops correlate with the average speed. We found that stops and the average speed in route A have a high linear correlation ($R^2=0.7553$). It is also intuitive that the number of stops in route A, which does not consist of highways but passes through streets in grid path structure, has a large effect on the average speed. Others have relatively low correlation coefficient. For example, the second largest correlation (with $R^2=0.2984$) is from stops and the average speed for route B. Brakes and the average speed have very small linear correlation.

\subsection{Analysis on Traces with Low Estimation Accuracy}
We analyzed the traces with low estimation accuracy (destination error more than 2 miles). For each of these traces, we analyzed the main cause of estimation mistakes. The factors misleading the algorithm can be grouped into following categories:
\begin{itemize}
\item \textbf{Homogeneity of roads:} If an area consists of many similar roads (e.g.~same speeds, similar intersection intervals, areas built in a grid), they cannot be distinguished.
\item \textbf{Unpredictable stops:} If there are several stops (due to traffic or constructions) not near any intersection, the speed trace will mislead our algorithm to find a wrong path.
\item \textbf{Very few stops:} A speed trace with very few stops lacks information about intersections for the ground truth, resulting many candidate paths cannot be effectively ranked.
\item \textbf{Traces with mostly slow speeds:} Without high speeds to rule out turns and constrain paths to a few major roads, the correct path is indistinguishable from any other. Road speed limit testing also does not help in this case.
\item \textbf{Limitations of OSM:} Nodes connected by the OSM are sometimes impossible to reach in real driving: connecting a road to a bridge over it, and entering non-driving paths (e.g.~railways, non-existing paths, private regions). 
\item \textbf{Unpredictable direction turnarounds:} Some drivers turned around their car in the midway of a path segment (e.g.~illegal u-turn, or drive through a place and then reverse direction).
\end{itemize}

Based on our analysis of low-accuracy traces from all three datasets, Table~\ref{tab:my-label} represents the number of traces that have estimation mistakes caused by each of the six factors. Although poor estimation of a trace may be due to several factors, we only labeled each trace based on the suspected root cause. For New Jersey dataset, limitations of OSM and traces with mostly slow speeds appear to be the top factors. For Seattle dataset, homogeneity of roads causes the most number of traces to have poor accuracy. Identical-route dataset consists of two routes with (route B) and without (route A) highway segment. For route A, unpredictable stops appears to be the main factor. However, for route B, homogeneity of roads is the main factor due to several highways near the same region: most of these traces end up in a different direction or on a different highway.

\begin{table}[]
\centering
\begin{tabular}{|c|c|c|c|c|}
\hline
\multirow{2}{*}{}     & \multirow{2}{*}{\begin{tabular}[c]{@{}c@{}}New Jersey \\ Dataset\end{tabular}} & \multirow{2}{*}{\begin{tabular}[c]{@{}c@{}}Seattle\\ Dataset\end{tabular}} & \multicolumn{2}{c|}{Identical-route Dataset} \\ \cline{4-5} 
                      &                                                                                &                                                                            & Route A               & Route B              \\ \hline
Homogeneity of roads  & 17                                                                             & 39                                                                         & 2                     & 18                   \\ \hline
Unpredictable stops   & 17                                                                             & 28                                                                         & 5                     & 6                    \\ \hline
Very few stops   & 13                                                                             & 21                                                                         & 1                     & 10                   \\ \hline
Mostly slow speeds    & 22                                                                             & 25                                                                         & 1                     & 3                    \\ \hline
OSM limitations       & 23                                                                             & 26                                                                         & 2                     & 3                    \\ \hline
Direction turnarounds & 3                                                                              & 3                                                                          & 0                     & 0                    \\ \hline
Total                 & 95                                                                             & 142                                                                        & 11                    & 40                   \\ \hline
\end{tabular}
\vspace{1pt}
\caption{Table shows number of traces with wrong estimation caused by each one of the six factors.
}
\label{tab:my-label}
\end{table}

\section{Discussion}
In this section, we discuss the definite version of our elastic pathing algorithm (accuracy and results), limitations, and importance of privacy-preserving data collection.

\textbf{Algorithm Accuracy and Results:}
Our definite version of elastic pathing algorithm estimates 17\% of traces from New Jersey dataset and 15.5\% traces from Seattle dataset with error within 250 meters. Our algorithm provides reliable estimation results for traces having good speed patterns. Such traces have no heavy traffic,
have very few random stops or maneuvers, and carry
enough information (e.g.~reasonable stops and speed values) to
distinguish with other roads.
Our algorithm can automatically prioritize those directions that seem to match well with the speed traces. Thus, it usually generates results instantly or within seconds for good speed traces. For ambiguous driving trips (e.g.~either because of homogeneous driving environments or poor speed patterns), it may take much longer and the estimation is usually not good.

Based on our results, OSM routing with shortest path does not improve the overall accuracy of our existing datasets.
However, it can be useful in estimating routes for a subset of drivers who mostly drive shortest routes in their daily commuting. Based on our results, routing method had different effects on suburban (New Jersey) and urban (Seattle) datasets. It also showed improvements on the accuracy for P4 and P6 in our New Jersey dataset and P13 in the Seattle dataset. 
In our case, the datasets are very comprehensive with various driving environments. The routing approach brings more penalty than benefits for our datasets.

On the other hand, using speed limits is important for our algorithm -- there is a large accuracy drop when removing this constraint for New Jersey dataset. However, removing speed limit constraint has very small impact for the Seattle dataset. This may be because the traces in Seattle dataset have similar speed limits. The types of roads (mainly consists of streets built in a grid) in urban areas are not as diverse as those in suburban areas. In this case, speed limits does not help in eliminating candidate paths.

Our algorithm accuracy highly depends on different driving habits as shown in our analysis for identical-route dataset. For route A and route B, our algorithm favors different behaviors (e.g.~different distribution patterns for stops and average speeds in Figure~\ref{fig:identical-route2} and Figure~\ref{fig:identical-route1}) due to different driving environments. However, they also favor some common driving features: both routes have high estimation accuracy when average braking deceleration is low, and the effect of braking has similar patterns.

\textbf{Limitations:}
One limitation is from the OSM routing. We applied OSM routing API currently available: finding the route with shortest distance. It does not have options to customize and find the fastest route to reach the destination for example. Intuitively, finding the fastest route is more challenging than finding the shortest route, as selection of fastest route depends on the traffic condition at the moment of driving.
Drivers may not necessarily take the shortest route, so a better routing API allowing different customization and settings (e.g.~finding the fastest route) may help improve our algorithm's estimation accuracy.

\textbf{Notes on Privacy-Preserving Data Collection:}
Our work is not directed against any company or organization. However, prior experience has shown that even well-intentioned uses of data can result in losses of privacy, and thus, we wish to highlight potential dangers of this
type of data collection. We do not claim that insurance companies
are violating policy holder privacy in this way. However,
the general principle of any privacy-preserving data collection
is to collect only the data that is necessary for a particular
application, and no more.

\section{Conclusions}
We presented our definite version of the elastic pathing algorithm which demonstrates an effective attack against user location privacy that uses only speed data from driving traces and an initial starting location. 
We have shown improvements over our initial work: 150 traces in total having destination error within 250 meters, comparing to 125 traces in our initial work, which is 20 percent improvement of 125 traces in this highly accurate estimation range. Overall, about half of driving traces (46\% for New Jersey dataset and 63\% for Seattle dataset) can be estimated with destination error within one mile of the ground truth. 37\% of all traces have destination error smaller than 0.5 mile.

We performed a comprehensive analysis on our algorithm, evaluated effect of applying speed limits, and compared how different driving behaviors can affect our algorithm with a new identical-route dataset. In addition, we built a tool that can visualize any driving trace with animation in the Google map. We showed how the OSM routing can be merged into our algorithm as an optional customization and explained when it can be useful and why it cannot be generalized for all drivers. We also analyzed traces with low estimation accuracy and summarized the cases that can mislead our algorithm. To the end, we have effectively shown that our algorithm can solve the major challenge of estimating the driving route using very limited data.

\ifCLASSOPTIONcompsoc

  \section*{Acknowledgments}
\else
  \section*{Acknowledgment}
\fi
This  material  is  based  upon  work  supported  by  the  National  Science  Foundation  under  Grant  Numbers  1228777 and 1211079. Xianyi Gao was supported by the National Science Foundation Graduate Research Fellowship Program under Grant Number 1433187. Any opinions, findings, and conclusions or recommendations expressed in this material are those of the authors and do not necessarily reflect the views of the National Science Foundation.

\ifCLASSOPTIONcaptionsoff
  \newpage
\fi

\bibliographystyle{IEEEtran}

\bibliography{ep-bibfile}
\balance

\end{document}